\documentclass[12pt,a4paper]{article}
\usepackage[utf8]{inputenc}
\usepackage[T1]{fontenc}
\usepackage{amsmath}
\usepackage{amsfonts}
\usepackage{amssymb}
\usepackage{graphicx}
\usepackage{natbib}
\usepackage{graphicx}
\usepackage{caption}
\usepackage{placeins}
\usepackage{bbm}
\usepackage{mathtools,xparse}
\usepackage{scalerel,stackengine}
\usepackage{float} 

\stackMath
\newcommand\reallywidehat[1]{%
\savestack{\tmpbox}{\stretchto{%
  \scaleto{%
    \scalerel*[\widthof{\ensuremath{#1}}]{\kern-.6pt\bigwedge\kern-.6pt}%
    {\rule[-\textheight/2]{1ex}{\textheight}}
  }{\textheight}%
}{0.5ex}}%
\stackon[1pt]{#1}{\tmpbox}%
}
\usepackage{geometry}
\usepackage{ctable} 
\usepackage[english]{babel}
\usepackage{blindtext}
\usepackage[affil-it]{authblk}
\usepackage[hidelinks,colorlinks=true,linkcolor=blue,citecolor=blue]{hyperref}
\title{\bf {Random Subspace Local Projections}\thanks{\href{mailto:Viet.Dinh@monash.edu}{Viet.Dinh@monash.edu}, \href{mailto:Didier.Nibbering@monash.edu}{Didier.Nibbering@monash.edu}, and \href{mailto:Benjamin.Wong@monash.edu}{Benjamin.Wong@monash.edu}.} \thanks{We thank the editor, three anonymous referees, Silvia Gon\c{c}alves, Eleonora Granziera, Luke Hartigan, \`{O}scar Jord\`{a}, Daniel Lewis, Adrian Pagan, Barbara Rossi and participants at seminars at the BI Norwegian Business School, 2023 Econometric Society of Australasia Meeting, 2023 International Conference in Computing in Economics and Finance,  the Advances in Local Projections and Empirical Methods for Central Banking Conference, Workshop of the Australasia Macroeconomic Society, OzMac Workshop, and the CAMA Workshop on Macroeconomic Challenges for Open Economies for helpful comments and suggestions. We acknowledge funding support from the Australian Research Council (DP190100202, DE200100693 and DP240100970).}}

\author{Viet Hoang Dinh}
\author{Didier Nibbering}
\author{Benjamin Wong}
\affil{Monash University, Australia}

\date{\today}
\begin{document}
\maketitle
\begin{abstract}
\noindent \footnotesize
We show how random subspace methods can be adapted to estimating local projections with many controls. Random subspace methods have their roots in the machine learning literature and are implemented by averaging over regressions estimated over different combinations of subsets of these controls. We document three key results: (i) Our approach can successfully recover the impulse response functions across Monte Carlo experiments representative of different macroeconomic settings and identification schemes. (ii) Our results suggest that random subspace methods are more accurate than other dimension reduction methods if the underlying large dataset has a factor structure similar to typical macroeconomic datasets such as FRED-MD. (iii) Our approach leads to differences in the estimated impulse response functions relative to benchmark methods when applied to two widely studied empirical applications. 
\end{abstract}
\thispagestyle{empty}
\newpage
\setcounter{page}{1}

\section{Introduction}\label{sec:introduction}
Impulse response functions (IRFs) are a common tool in macroeconomics to study the dynamics of variables in response to shocks. A recent development is to estimate IRFs using local projections (LPs), developed by \cite{jorda2005estimation} \citep[e.g., see][]{tenreyro2016pushing,ramey18state,swanson2020measuring}. LPs are single-equation methods, and can be thought of as the direct forecast counterpart of traditional multiple equation systems, such as a vector autoregressions (VARs), which rely on iterating on a system of equations. 

A parallel development in the macroeconometrics literature has been the use of large datasets. The cost of putting together a large dataset for empirical analysis, at least for U.S. macroeconomic data, is now becoming exceedingly trivial, especially with datasets such as FRED-MD and FRED-QD \citep{mccracken2020fred}. Like any regression model, the choice of control variables to use in an LP forms part of the model specification. With the availability of large datasets, the dimension of this control set has become potentially large.

The contribution of our paper is to introduce random subspace methods for the estimation of IRFs using LPs. The method is simple to implement by three steps: First, we include a random subset of controls in the LP and estimate the IRF. Second, we repeat the first step many times. Finally, we take the average IRF. Through averaging over random subsets of controls, where the subsets are generated independently of the data, the variance of the IRF estimate is reduced while maintaining most of the signal of the controls. 

A practical implication of our approach is that it frees up the researcher to focus on the IRFs, which is the key object of interest.
Conditional on a base set of controls ensuring identification of the IRF, the random subspace method deals with the issue of appropriately including controls to improve estimation efficiency. In case a large set of controls is required for identification, the averaging across different subspaces mitigates the risk of identification bias, while avoiding the signal of the shock of interest to get swamped by the controls. 
This is an appealing practical feature because the coefficients on the controls are almost never the object of interest, as long as the appropriate controls are included in order to obtain accurate estimates of the IRFs.

We highlight that random subspace methods apply more naturally within the LP setting despite the now-known result that LPs and VARs recover the same IRF in population \cite[see][]{plagborg2021local}. In a VAR setting, an additional control variable implies an extra equation. In the LP setting, extra variables imply just more controls within the single equation. With the potential control variables numbering into the tens or hundreds, appropriately dealing with extra, possibly extraneous, controls is arguably more manageable than estimating tens or hundreds of extra, possibly irrelevant, equations.

Our key results are as follows. First, we show that random subspace methods can recover the true IRFs under plausible factor structures and usual identification techniques for the structural shock of interest. We base this conclusion of two sets of Monte Carlo experiments,  one using a real business cycle model with fiscal foresight introduced by \cite{leeper2013fiscal}, and one using the dynamic factor model of \cite{stock2016dynamic}. The proposed method is implemented with both SVAR identification \citep{plagborg2021local}, and instrumental variable (IV) identification \citep{stock2018identification}.
Random subspace methods help accurately estimate IRFS using a large set of controls that exhibits different factor structures resembling macroeconomic datasets. 

Second, the Monte Carlo experiments demonstrate that random subspace LPs are competitive in terms of mean squared error to a wide range of methods, including commonly used VAR methods and LPs with alternative dimension reduction methods. More specifically, the use of random subspaces improves upon the use of factors when the factor structure resembles that of U.S. macroeconomic datasets such as FRED-MD. Random subspaces are also favoured over variable selection methods, possibly in part due to the dense structure of macroeconomic data \cite[see][]{giannone2021economic}, though whether our method outperforms VAR based methods can depend on the specific context.

Third, we document in two widely studied empirical applications (i.e. the effect of the macroeconomy to an identified monetary policy and technology shock) that random subspace LPs produce meaningful differences in the estimated IRFs relative to common empirical strategies. Our results are encouraging as it suggests that the random subspace LP is a simple procedure or robustness check for practitioners using LPs who may be concerned that they may have omitted relevant controls. 

We link our work to two strands of the broader literature. First, although random subspace methods have their roots in the machine learning literature, researchers have recently applied them to improve forecast accuracy for economic indicators based on a large number of possible predictors \citep{koop2019bayesian,kotchoni2019macroeconomic,boot2019forecasting,pick2022multi}. In general, they find strong performance of random subspace methods across different macroeconomic forecasting exercises. Since the datasets used in macroeconomic structural analysis are similar, exploring random subspace methods to estimating IRFs seems natural. 

Second, various approaches
have been proposed to improve the efficiency of the LP estimator \cite[e.g.][]{barnichon19smooth,ferreira2023bayesian,lusompa2023local,ho2023averaging}, which is known to have a lower bias but higher variance than VARs \cite[see, e.g.][]{li2022local}. One can view our approach in a similar vein. Given that proliferation of controls leads to inefficiency, researchers naturally economize on the number of controls. Random subspace methods are a form of regularization, in which averaging across subsets of controls targets the bias-variance trade-off between omitting potentially relevant information and including all available data as controls. 

The remainder of the paper is organised as follows: Section~\ref{sec:lpcontrols} provides a detailed discussion of our framework. Section \ref{sec:cookbook} summarizes how to implement our procedure in a step-by-step manner.
Section~\ref{sec:MCsimullongrunexample} uses a Monte Carlo exercise to understand how random subspace methods help in appropriately estimating IRFs. Section~\ref{sec:empapplications} applies our proposed approach to two widely studied empirical applications. Section~\ref{sec:Conclusion} provides some concluding remarks. 
 
\section{Method}\label{sec:lpcontrols}
Consider the standard LP regressions one specifies to estimate the IRF to the variable $y$ from an exogenous one unit impulse on $x_t$:
\begin{equation}\label{eq:lp_base}
 {y}_{t+h} = \mu_h+ \beta_h x_t+ {\Phi}_{h}'{W}_t +\xi_{t+h} ,\quad h=0,1,\dots,H,
\end{equation} 
where $\mu_h, \beta_h$, and ${\Phi}_h$ are projection coefficients, $\xi_{t+h}$ is the projection error, and ${W}_t$ is a vector of controls. 
We are interested in the response of $y_{t+h}$ with respect to an exogenous one unit impulse to $x_t$, which equals $\beta_{h}$. The problem we investigate in this paper is how one deals with the control set ${W}_t$. While one is almost never interested in the coefficients ${\Phi}_{h}$, the specification of the control set can matter to obtaining accurate estimates of $\beta_{h}$. 

We briefly highlight two roles the controls play in the estimation of $\beta_{h}$. First, the set of control variables accounts for relevant determinants which are correlated with $x_t$. In this setting, failure to include the relevant controls results in omitted variable bias, and so biased estimates of the IRF. Second, even if $x_t$ is strictly exogenous, or we possess a strictly exogenous instrument, including additional controls in the projection may reduce the variation in forecasting $y_{t+h}$, which may result in a more precise estimate for $\beta_{h}$. However, in this setting, the effect is less obvious. If the included set of controls is too large, the increase in parameter uncertainty reverses possible efficiency gains, and may then increase the variance of the estimate for $\beta_{h}$.

With large macroeconomic datasets like FRED-MD, empirical work in macroeconomics now has access to over a hundred variables that may serve as potential controls. This leads to the familiar trade-off for practitioners where omitting relevant controls results in bias, but including irrelevant variables leads to an increase in variance. In cases where a variable is known to be irrelevant, such a variable should be omitted, although in most empirical settings it is not obvious whether a variable is irrelevant or not. This is the setting that we propose solving through random subspace local projections (RSLP): we alleviate the issue of specifying ${W}_t$, mindful that a practitioner is often only interested in $\beta_{h}$.

\subsection{Random Subspace Local Projections (RSLP)}\label{sec:rslp}
Instead of a generic set of controls ${W}_t$, we make a distinction between the $p_V\times 1$ vector ${V}_{t}$ of variables which are considered essential controls, and the $p_G\times 1$ vector ${G}_t$ of possibly relevant controls. Rewriting \eqref{eq:lp_base} in terms of these essential and possible controls results in
\begin{align}\label{eq:lpwolf1}
 {y}_{t+h}&= \mu_h+ \beta_h x_t+ {\Theta}_{h} {V}_{t}+{\Psi}_h {G}_t+\xi_{t+h} ,
\end{align}
where ${\Theta}_{h}$ and ${\Psi}_h$ are the projection coefficients of ${V}_t$ and ${G}_t$, respectively. While nothing precludes specifying ${V}_t$ as an empty set, most macroeconomic applications would \textit{a priori} treat some variables as essential. For example, the usual monetary policy shock application would include at least some measure of real activity, inflation, and the interest rate. The identification strategy for $\beta_h$ may also inform which variables have to be included in ${V}_t$, a matter which we discuss in detail in Section~\ref{sec:identifications}.\footnote{Even in cases where one possesses a strictly exogenous and valid instrument --and no controls are required for the LP estimator to be consistent-- the inclusion of controls known to be relevant in ${V}_t$ may improve the accuracy of the estimate for $\beta_h$.} 

Some applications may only provide essential control categories, but leave the precise variable unspecified. For example, even if the model should feature an interest rate variable, there are many such variables in datasets like FRED-MD. Our empirical examples select one of the interest variables to be included in $V_t$ in this case, and include the rest in ${G}_t$. Alternatively, Appendix~\ref{A:categories} extends our approach to using prior knowledge on which categories are essential, without having to select a specific variable from those categories.

The number of possibly relevant controls in ${G}_t$ is potentially large as previously motivated. Instead of estimating $\beta_{h}$ in \eqref{eq:lpwolf1}, a form of dimension reduction is usually applied to ${G}_t$. Consider the linear projection from \eqref{eq:lpwolf1} with a $k \times p_G$ compression matrix ${R}^{(j)}$ indexed by $j$, where $k\le p_G$:
\begin{align}\label{eq:rslp} 
{y}_{t+h}=\mu_h^{(j)}+ \beta_h^{(j)}x_t+{\Theta}_{h}^{(j)} {V}_{t}+ {\Gamma}_{h}^{(j)}{R}^{(j)}{G}_t+\xi_{t+h}^{(j)},
\end{align}
where ${\Gamma}_{h}^{(j)}$ is a $k$-dimensional vector of projection coefficients, instead of the $p_G$-dimensional vector of projection coefficients ${\Psi}_h$ in \eqref{eq:lpwolf1}. The construction of the compression matrix can be data-driven. For instance, variable selection methods use data to estimate a selection matrix for ${R}^{(j)}$. Factor-augmented models take ${R}^{(j)}$ as the matrix of the principal component loadings corresponding to the $k$ largest eigenvalues from the sample covariance matrix of ${G}_t$.

The random subspace approach to dimension reduction is to generate the elements of ${R}^{(j)}$ from a probability distribution that is independent of the data. More precisely, $R^{(j)}$ is sampled from a uniform distribution across all combinations of the $p_G$ available predictors of size $k$. It follows that ${R}^{(j)}$ is a random subset selection matrix which randomly selects a subset of $k$ predictors out of the $p_G$ predictors. For instance, if there are 5 possible predictors and we wanted to choose 3, $p_G=5$ and $k=3$, and one possible draw $j$ could be the following:
\begin{align*}
{R}^{(j)} = 
\begin{bmatrix}
	0 & 1 & 0 & 0 & 0\\
	0 & 0 & 1 & 0 & 0\\
    0 & 0 & 0 & 0 & 1
\end{bmatrix}.
\end{align*}

Conditional on a draw ${R}^{(j)}$, $\beta_h^{(j)}$ in \eqref{eq:rslp} can be estimated using least squares. The random subspace estimate for $\beta_h^{(j)}$ is constructed by averaging over the least squares estimates $\hat{\beta}_h^{(j)}$ corresponding to different draws ${R}^{(j)}$, with $j=1,\dots,n_R$:
\begin{align}
	\hat{\beta}_{h}=\frac{1}{n_R}\sum_{j=1}^{n_R}\hat{\beta}_{h}^{(j)}. \label{eq:irfrslp}
\end{align}
Through averaging, random subspace methods reduce the variance of the LP estimate, while it maintains most of the signal of the controls. We note that one could combine these subspaces using weights reflecting the performance of each subspace, instead of equal weights. However, estimation of these optimal weights introduces additional variance, and hence a simple average is generally hard to beat, as shown by \cite{elliott2013handbook}.\footnote{In Appendix~\ref{A:weights}, we use a weighting scheme based on the Bayesian information criterion as discussed by \cite{elliott2013complete} in our empirical examples, which indeed results in little differences relative to equal weighting.} 

Implementing the random subspace LPs requires a choice for the subspace dimension $k$. We find that setting $k=50$ is a reasonable choice in our Monte Carlo experiments and empirical applications. While one could use an information criterion to choose the dimension, as we discuss in Appendix~\ref{A:selection}, this may not work well in macroeconomic settings with a large number of controls and a small sample size. Sensitivity analyses reported in Appendix~\ref{A:dimension_monte_carlo} suggest that the optimal subspace dimension is often in the 40-60 range. 
Hence, we proceed with an empirical strategy that uses $k=50$, and subsequently check for the robustness of the results.

Random subspace methods have shown to improve forecast accuracy for economic indicators based on a large number of possible predictors \citep{koop2019bayesian,kotchoni2019macroeconomic,boot2019forecasting,pick2022multi}. 
The relative effectiveness of the random subspace approach may be explained by three features of macroeconomic data. First, macroeconomic variables are often highly correlated. Variables that are irrelevant when all relevant controls are included, are often correlated with omitted relevant controls in a random subset. Hence, the use of random subsets results in a small bias while it reduces the variance. Second, most macroeconomic data is characterized by a low frequency and low signal. As a result, the selection of the controls in ${G}_t$, or the derived factors, may be subject to substantial uncertainty. Third, \citet{giannone2021economic}, among others, find that often many variables are relevant in explaining macroeconomic quantities of interest. This makes it even more challenging to find an accurate low-dimensional representation of the data by variable selection methods. \citet{boot2020subspace} discuss random subspace methods and its theoretical justification in the context of macroeconomic forecasting in more detail.

\subsection{Structural identification within RSLP}\label{sec:identifications}

The exposition so far discusses how one can use random subspace methods to estimate IRFs in an LP setting. We now connect our approach to recovering the object of interest, namely the IRF from an underlying structural model.

Define ${w}_t$ as an $n \times 1$ vector of macroeconomic variables, in which both the variable of interest $y_t$ and the variable $x_t$ to which we are introducing an exogenous impulse in \eqref{eq:lp_base} are included. The variables in ${w}_t$ are driven by a $m \times 1$ vector of independent and identically distributed (i.i.d.) shocks ${\epsilon}_t$:
\begin{equation}
\label{eq:VMA}
{w}_t = {\Theta}(L) {\epsilon}_t, \qquad {\epsilon}_t \sim i.i.d. (0,{I_m}),
\end{equation}
with lag polynomial ${\Theta}(L) \equiv \sum_{j=0}^{\infty} {\Theta}_j L^j$, and $n \times m$ coefficient matrices ${\Theta}_j$.

Suppose, without loss of generality, that we are interested in the effect of the first shock $\epsilon_{1t}$, and $y_t$ and $x_t$ are respectively the $i^{th}$ and $j^{th}$ variable in the vector ${w}_t$. After normalizing $\epsilon_{1t}$ to imply a unit increase in $x_t$,\footnote{This normalization is without loss of generality since shocks are unobserved, so the variance and sign of shocks are ultimately normalized. Our choice of normalization simplifies the exposition from needing to introduce a parameter to link $\epsilon_{1t}$ to $x_t$.} we can write $x_t$ as a linear combination of shocks:
\begin{align} \label{eq:x1}
x_t=\epsilon_{1t}+ f(\epsilon_{2:m, t}, \epsilon_{t-1},\epsilon_{t-2},...),
\end{align}
where $\epsilon_{2:m, t}\equiv(\epsilon_{2t},...,\epsilon_{mt})'$ and $f(.)$ is a linear combination of its argument. Leading the $i^{th}$ equation of \eqref{eq:VMA}, which describes the dynamics of $y_t$, by $h$ periods, and subsequently rearranging \eqref{eq:x1} to substitute out for $\epsilon_{1t}$, we obtain
\begin{equation}
\label{eq:x2}
   y_{t+h} = \theta^{i1}_h x_t + f(\epsilon_{t+h} ...,\epsilon_{t+1}, \epsilon_{2:m, t}, \epsilon_{t-1},\epsilon_{t-2},...),
\end{equation}
where the argument in $f(\cdot)$ now subsumes $y_{t+h}$ being a linear function of past and future shocks and $f(\epsilon_{2:m, t}, \epsilon_{t-1},\epsilon_{t-2},...)$ from \eqref{eq:x1}. The object of interest is $\theta^{i1}_h$, as it is the IRF from \eqref{eq:VMA}. Based on the form in which \eqref{eq:x2} is written it would be analogous to $\beta_h$ in \eqref{eq:lp_base}. Due to endogeneity, \eqref{eq:x2} cannot be consistently estimated since both $x_t$ and $y_{t+h}$ are a function of {all} past and current shocks. More precisely, in our context, $x_t$ is not only correlated with $\epsilon_{1t}$ but also with other shocks that affect $y_{t+h}$. 

\subsubsection{Identification through instruments or external variation}

The IRF can be estimated using two-stage least squares and an instrument $z_t$ that satisfies the following conditions, termed Condition LP-IV by \cite{stock2018identification}:
\begin{enumerate}
    \item{$\mathbb{E}\left[\epsilon_{1t}z_t \right] \neq 0$ (relevance)},
    \item{$\mathbb{E}\left[\epsilon_{2:m, t}z_t \right] = {0}$ (contemporaneous exogeneity)},
    \item{$\mathbb{E}\left[\epsilon_{t+j}z_t \right] = {0},\, j \neq 0$ (lead-lag exogeneity)}.
\end{enumerate}

With a strictly exogenous instrument satisfying the LP-IV conditions, the instrument and the shocks are assumed to be uncorrelated contemporaneously and at all lags, and hence the IRF can be estimated consistently. Directly observing the shock is equivalent to observing a $z_t$ which is perfectly correlated with $\epsilon_{1t}$. In this case, Condition LP-IV allows one to directly regress the shock on $y_{t+h}$, similar as in approaches such as \cite{romer2004new}. In these settings, the inclusion of controls is to the extent that they may lead to efficiency in finite samples, but they are not required for the identification of the IRF.

Condition LP-IV is a strong assumption but can be relaxed by including controls:
\begin{eqnarray}
\mathbb{E}\left[\epsilon_{2:m_\epsilon, t}^{\perp} z_t^{\perp} \right] &=& {0},\, j \neq 0 \, \text{(conditional contemporaneous exogeneity)}, \label{eq:cce}\\
    \mathbb{E}\left[\epsilon_{t+j}^{\perp} z_t^{\perp} \right] &=& {0},\, j \neq 0 \, \text{(conditional lead-lag exogeneity)}, \label{eq:cle}
\end{eqnarray}
where $u_t^{\perp} = u_t - Proj(u_t\mid {W}_t)$ for some variable $u$. This is a more relevant setting empirically, given widely used instruments for monetary policy shocks have been shown to be either contaminated by other shocks \cite[see][for a prominent example]{miranda2018transmission}, or possibly forecastable by other macro variables \cite[see][]{ramey2016macroeconomic}. These suggest violations of strict exogeneity. Controls that account for this information can aid in satisfying conditional exogeneity, and so help constructing a valid instrument $z_t^{\perp}$.\footnote{This argument is analogous to what \cite{lloyd_manuel2401} label as the one-step approach, where one uses controls in the first stage to render the instrument conditionally exogenous.} If there are particular controls that one knows to help the instrument to satisfy conditional exogeneity, they should be treated as essential controls. In addition,  RSLP may efficiently account for the information in the possibly relevant controls so that the conditions in \eqref{eq:cce} and \eqref{eq:cle} are satisfied.

We thus extend random subspace methods to the two-stage least squares settings where the first and second stage regressions are
 \begin{align} 
 {x}_{t}&=\alpha^{(j)}+ \rho^{(j)} z_t + {\Lambda}^{(j)} {V}_t+{\Upsilon}^{(j)}{R}^{(j)}{G}_{t}+\eta_t^{(j)},   \label{eq:rslpiv1}\\
 		y_{t+h}&=\mu_h^{(j)}+ \beta_{h}^{(j)} \hat{x}_{t}^{(j)}+{\Theta}_{h}^{(j)}{V}_t+{\Psi}_{h}^{(j)}{R}^{(j)}{G}_{t}+\xi_{t+h}^{(j)},  \label{eq:rslpiv2}
 \end{align}
where ${\Lambda}^{(j)}$ and ${\Upsilon}^{(j)}$ are the projection coefficients from the first stage regression, $\hat{x}_t$ is the fitted value from \eqref{eq:rslpiv1}, and the estimates for $\beta_{h}^{(j)}$ in \eqref{eq:rslpiv2} are used in \eqref{eq:irfrslp} to average across random draws for selection matrices.

\subsubsection{Implementing SVAR identification}

\cite{plagborg2021local} show that SVAR or system-based identification (i.e. recursive identification, long-run restrictions, and sign restrictions) can be implemented using LPs. These identification schemes require SVAR invertibility, which allows one to map a VAR in $w_t$ to the form in \eqref{eq:VMA} by inverting the VAR lag polynomials. Under invertibility there is a linear combination of $w_t$ and its lags which accounts for $f(\epsilon_{2:m, t}, \epsilon_{t-1},\epsilon_{t-2},...)$ in \eqref{eq:x1}, so that what gets used in place of $x_t$ in \eqref{eq:x2} identifies the shock of interest $\epsilon_{1t}$. If not all relevant information is included, identification fails due to not satisfying invertibility. The literature has also long recognized the link between invertibility and the inclusion of all relevant information \cite[see, e.g.][]{hansen2019two,fernandez2007abcs,stock2018identification}.

Where a standard LP has a VAR representation estimating the same IRF in population, estimating an IRF by RSLP is analogous to averaging over the IRFs from VARs with different sets of variables, but the same identification strategy.\footnote{Note that we will be applying the same sign, long, or short run restriction across different subspaces. Therefore, all that is changing is the information set, or the set of controls, one is using to apply the identification scheme. In this regard, the approach has connections to the approach of \cite{ho2023averaging}, although they allow the identifying restrictions to differ across the different models that they average over.} 
In cases where the essential controls in $V_t$ are sufficient for invertibility, RSLP may help reducing the variance in the IRF estimates by regularizing over the information from the rest of the dataset. In cases where the possibly relevant controls in $G_t$ are required to satisfy invertibility, and thus play a role with identification, the averaging over different subspaces of $G_t$ implicitly takes uncertainty about the exact identifying assumptions into account. The averaging over many different information sets is also related to the estimation of IRFs by factor models or large Bayesian VARs, such as respectively those by \cite{bernanke2005measuring} and \cite{banbura2008large}. These methods may be viewed as alternative approaches to fulfill the invertibility condition by applying dimension reduction to large information sets.

To illustrate how SVAR identification may be implemented within an RSLP, we consider a recursive identification scheme from the seminal work by \citet{christiano1999monetary}. They identify a monetary policy shock by restricting the contemporaneous effect of ``fast moving'' variables on the Federal funds rate to be zero, and the contemporaneous effect of monetary policy shocks on ``slow moving'' variables to be zero. This suggests that contemporanous values of the ``slow moving'' variables, such as real activity and price variables, have to be included in the LP. The number of available real activity and price variables in modern macroeconomic datasets is large, and the identification scheme does not provide guidance on which specific variables to select.\footnote{For example, it is known that researchers have used industrial production, unemployment, employment, or real GDP, amongst others, to reflect real activity. While the motivation is to include at least one real activity variable, there is little guidance of which one to include. From this perspective, one could argue that researchers aim to control for the same information even if they disagree on which precise variable reflects that information.
} By including a small number of variables in $V_t$, such as industrial production and inflation, and the remaining available ``slow moving'' variables in $G_t$, one can view RSLP as a procedure to balance the risk of non-invertibility with overfitting.

\subsection{Error bands with RSLP} \label{sec:error_bands}

We briefly discuss how to construct error bands for the IRF estimated by RSLP. A key challenge is that the random subspace literature, like most machine learning methods, provides little guidance on how to quantify estimation uncertainty.

The standard error of a standard LP can be estimated with a block bootstrap to account for serially correlated errors \cite[see, e.g.][]{kilian2011reliable,inoue2023significance}.\footnote{We note that if one were certain that the lags of the correct variables are included as essential controls, it may be possible to obviate the need to account for serial correlation \cite[see][]{montiel2021local} and just consider the cross correlation across the regressions.}  Since RSLP is constructed by averaging across LP regressions, the bootstrap should not only preserve correlation across residuals in each regression, but also the correlation across the regressions. Therefore, a natural choice would be a block bootstrap, using the same block of residuals across regressions to also preserve the correlation across subspace estimates.
Although estimating a single RSLP requires $n_R$ least squares regressions, which would at most take a few seconds in a modern macro context, the bootstrap may amount to substantial computational costs. Given each horizon is bootstrapped separately, an IRF for one variable would require bootstrapping $n_R \times (H+1)$ regressions. The computation time of this procedure can be alleviated by parallelizing the computation across subspaces.

Since the block bootstrap for RSLP is computationally expensive, we also discuss an alternative approach derived from \citet{buckland1997model}. This approach relies on the assumption that there is perfect correlation across the IRF estimates from different subspaces.  Although the correlation across subspaces is high in practice, this assumption is likely too strict, and therefore the bands will be conservative. The error bands of standard LP bands are already known to be wide, although RSLP may alleviate this issue by efficiently incorporating controls that help explain variation in the variable of interest. The implementation details of both the bootstrap and the \citet{buckland1997model} approach are deferred to Appendix~\ref{sec:RSLPsd}.

\section{Implementing RSLP}\label{sec:cookbook}
We briefly summarize the preceding discussion to outline how RSLP can be implemented:

\begin{enumerate}
    \item{Make a distinction between essential controls, $V_t$, and the  controls that the random subspace method will be applied to, $G_t$. This distinction specifies the number of possible covariates that each random subspace matrix will select from, $p_G$. }
    \item{Choose a subspace dimension, $k$. The subspace dimension dictates how many covariates are included in each subspace regression. In practice, we find a subspace dimension of 50 to work well with U.S. macroeconomic data.}
    \item{Randomly draw from the $p_G$ possible controls, assigning equal probability that any of these $p_G$ controls can be chosen. Create a matrix indexed by $j$, $R^{(j)}$ of dimension $k \times P_G$, which acts as a selector matrix that sets the appropriate element to 1 if a control is chosen, and zero otherwise.}
    \item{For each horizon $h \in \lbrace 0, 1, \hdots, H \rbrace $, estimate ${y}_{t+h}=\mu_h^{(j)}+ \beta_h^{(j)}x_t+{\Theta}_{h}^{(j)} {V}_{t}+ {\Gamma}_{h}^{(j)}{R}^{(j)}{G}_t+\xi_{t+h}^{(j)}$. Depending on the identification strategy, $x_t$ may need to be constructed with a first stage regression. This regression may be similar to \eqref{eq:rslpiv1} if one used an external instrument, or is specified using system-based identification restrictions such as long-run or sign restrictions.\footnote{See \cite{plagborg2021local} for a general discussion on how to implement system-based identification in an LP setting. \cite{alpanbda2021} discuss sign restrictions, and Appendix \ref{sec:SVARidentification} long-run restrictions in the spirit of \cite{blanchard1988dynamic} in LPs.} Because the first stage regression also requires choosing controls, the $R^{(j)}$'s will also be used in the first stage.}
    \item{Repeat steps 3 and 4 a large number of times. In practice, we find a number of subspace draws $n_R = 1000$ to be sufficiently large that results are stable. The estimated IRF for variable $y$ at horizon $h$ to the identified shock of interest, can be obtained via averaging over the $n_R$ draws:  $\beta_h = \frac{1}{n_R}\sum_{j=1}^{n_R} \beta_h^{(j)}$.}
\end{enumerate}

\section{Monte Carlo Experiments}\label{sec:MCsimullongrunexample}
As an illustration of the RSLP procedure, we consider Monte Carlo experiments based on the real business cycle model with fiscal foresight by \cite{leeper2013fiscal}. The phenomenon of fiscal foresight occurs when at time $t$ agents know the tax rate they will face at time $t+h$. The model includes income taxes, inelastic labor supply, and full capital depreciation. The log-linearized equilibrium condition for capital is given by\footnote{See \cite{leeper2013fiscal}, page 1118, Equation (4).}
\begin{align}
	k_t=\alpha k_{t-1}+u_{at}-(1-\theta)\frac{\tau}{1-\tau}\sum_{k=0}^{\infty}\theta^{k}E_t\hat{\tau}_{t+k+1}, \label{eq:equicap}
\end{align}
where $k_t$ and $\hat{\tau}_t$ are respectively the percentage deviations from the steady state capital and tax rate, $\tau$ is the steady state value of the tax rate, $\theta$ and $\alpha$ are parameters satisfying the inequalities $0<\theta<1$ and $0<\alpha<1$, and $u_{at}$ is an i.i.d. technology shock. The tax rule is $\hat{\tau}_{t+h}=u_{\tau t}$, where $u_{\tau t}$ is an i.i.d. tax shock. Allowing $h=2$ (a two-period foresight), \eqref{eq:equicap} becomes 
\begin{align}
	k_t=\alpha k_{t-1}+u_{at}-\kappa(\theta u_{\tau t}+u_{\tau t-1}), \label{eq:equicapt2}
\end{align}
where $\kappa=\tau(1-\theta)/(1-\tau)$. The structural moving average representation equals 
\begin{align}
\begin{bmatrix}
	\hat{\tau}_t\\
	 k_t
\end{bmatrix}
=\begin{bmatrix}
	L^2 & 0\\
	-\frac{\kappa(L+\theta)}{1-\alpha L} & \frac{1}{1-\alpha L}
\end{bmatrix}
\begin{bmatrix}
	u_{\tau t}\\
	u_{at}
\end{bmatrix}. \label{eq:fiscalforesightDGP}
\end{align}
To parametrize the above, we follow \cite{leeper2013fiscal} and \cite{forni2014sufficient} by setting $\theta=0.2673$, $\tau=0.25$, $\alpha=0.36$, and $u_{\tau t}$ and $u_{at}$ to be from independent standard Normal distributions.

\subsection{Information sets and identification schemes}\label{sec:DGP}
The econometrician's objective is to estimate the IRFs of both taxes and capital to a tax shock. Due to fiscal foresight, there is insufficient information to recover the IRFs by just observing capital and the tax rate at time $t$. The Monte Carlo experiments consider different ways of extending the information set of the econometrician.

\paragraph{Strictly exogenous instrument} First, we allow the econometrician to observe a strictly exogenous instrument together with additional series that can aid in recovering the IRFs. The instrument $z_t$ for the tax shock $u_{\tau t}$ is generated as
\begin{align}\label{eq:z_exo}
    z_t = 0.7 u_{\tau t} +\nu_{1t-1} +\nu_{2t-1} + \epsilon_{z}, \quad  \epsilon_{z} \sim N(0,0.01),
\end{align}
where the shocks $\nu_{1,t}$ and $\nu_{2,t}$ are both generated from a $N(0,4)$ independent from the structural model. The instrument identifies the IRF without controls. However, being able to account for variation in $z_t$ due to $\nu_{1t-1}$ and $\nu_{2t-1}$ can lead to efficiency gains. To this end, 100 informational variables $y_{it}^*$ are generated as 
\begin{align}
	\label{eq:inforseries_noise}
		y_{it}^* = b_i \nu_{1t} +(1-b_i) \nu_{2t}+\xi_{it}, \quad  \xi_{it} \sim N(0,\sigma_i^2), \quad i=1,\dots, 100,
\end{align}
where $b_i$ is a Bernoulli random variable assuming value 1 with probability 0.1. 

\paragraph{Conditionally exogenous instrument} Second, we allow the econometrician to observe an instrument for the tax shock which is only exogenous conditional on a set of informational variables.  The instrument $z_t$ is generated as
\begin{align}\label{eq:z_endo}
    z_t = 0.7 u_{\tau t} +u_{at-1} + u_{\tau t-1} + \epsilon_{z}, \quad  \epsilon_{z} \sim N(0,0.01),
\end{align}
where $u_{a t-1}$ and $u_{\tau t-1}$ are structural shocks. Failure to control for these lagged shocks will lead to a violation of lead-lag exogeneity. To be able to recover the IRFs with this instrument, we generate 100 informational variables $y_{it}^*$ as
\begin{align}
	\label{eq:inforseriesDGP}
		y_{it}^* = b_i u_{\tau t-1} +(1-b_i) u_{a t-1}+\xi_{it}, \quad  \xi_{it} \sim N(0,\sigma_i^2), \quad i=1,\dots, 100.
\end{align}
The settings with strictly and conditionally exogenous instruments will help to understand the extent to which RSLP can achieve either reduction in estimation variance or identification bias.

\paragraph{SVAR identification} Third, we assume that the econometrician knows that the technology shock has no cumulative effect on the tax rate, and has access to the informational series from \eqref{eq:inforseriesDGP}. As shown by \cite{leeper2013fiscal}, the model implied by \eqref{eq:fiscalforesightDGP} does not admit an invertible SVAR representation in taxes and capital as the model is missing the two period ahead tax rate (or the current tax shock) in the information set. Since the informational variables contain this information, including these variables admits an invertible VAR representation. It follows from \cite{plagborg2021local} that subsequently applying SVAR identification with this additional information in an LP can recover the IRF.\footnote{We note that \cite{forni2014sufficient} already demonstrate that one can use SVAR identification to recover the IRF in this setting by using a FAVAR to account for this information. We impose the SVAR identification restriction in an LP setting by using the procedure presented by \cite{plagborg2021local}. We leave the implementation details to Appendix~\ref{sec:SVARidentification}.}

\paragraph{Controls with a strong and weakened factor structure}
 To examine the role of the strength of the factor structure of the information series in recovering the IRFs, we consider two settings that differentiate the factor structure of the informational variables in \eqref{eq:inforseries_noise} and \eqref{eq:inforseriesDGP}:
\begin{eqnarray}\label{eq:inforseriesstrong_weak}
    \text{Strong case:}& \sigma_i \sim U(0,1),\quad
    \text{Weak case:}& \sigma_i \sim U(0,4).
\end{eqnarray}
The strong case mimics a structure that one would probably encounter in a panel where these individual series have a high level of comovement, such as cross country asset prices and interest rates \cite[e.g.][]{miranda2020us}, or commodity prices \cite[e.g.][]{west2014factor,alquist2020commodity}. 
The weak case is akin to the sort one would expect to encounter when using U.S. macroeconomic data: Appendix~\ref{sec:datastructure} shows that the first two factors of both the FRED-MD and FRED-QD dataset explain a similar amount of variation as in the weak case.

We run each of the three identification schemes with both the strong and weak cases. This amounts to six Monte Carlo experiments, each simulating 1000 artificial datasets of 200 observations for capital and the tax rate according to \eqref{eq:fiscalforesightDGP}. The experiments corresponding to different information sets and identification schemes accompany each of these artificial datasets with: an exogenous instrument and informational variables generated from \eqref{eq:z_exo} and \eqref{eq:inforseries_noise}; a conditionally exogenous instrument and a set of informational variables from \eqref{eq:z_endo} and \eqref{eq:inforseriesDGP}; informational variables from \eqref{eq:inforseriesDGP} within SVAR identification.

\subsection{RSLP can recover the true impulse response functions}

To understand if RSLP can recover the IRF from the DGP, we study the expectation of the estimated IRFs. For each artifical dataset, RSLP constructs the IRF following the procedure in Section~\ref{sec:rslp} with 1000 draws of the selection matrix ${R}^{(j)}$ and ${G}_t$ equal to the first lag of $y^*_t$. To investigate whether RSLP is a valid approach to accounting for the omitted information, we estimate LPs without ${G}_t$, which we label LP with a base set of controls. Both specifications include the two lags of tax rate and capital in ${V}_t$, together with two lags of the instrument when using IV identification. Appendix~\ref{sec:SVARidentification} elaborates on the specification of the LPs when using SVAR identification. 

\begin{figure}[t!]
\caption{Impulse response functions in the Monte Carlo experiments} \vspace{-5mm}
\begin{center}
 (a) Strictly exogenous instrument
\includegraphics[width=\textwidth,trim={10 0 10 0}]{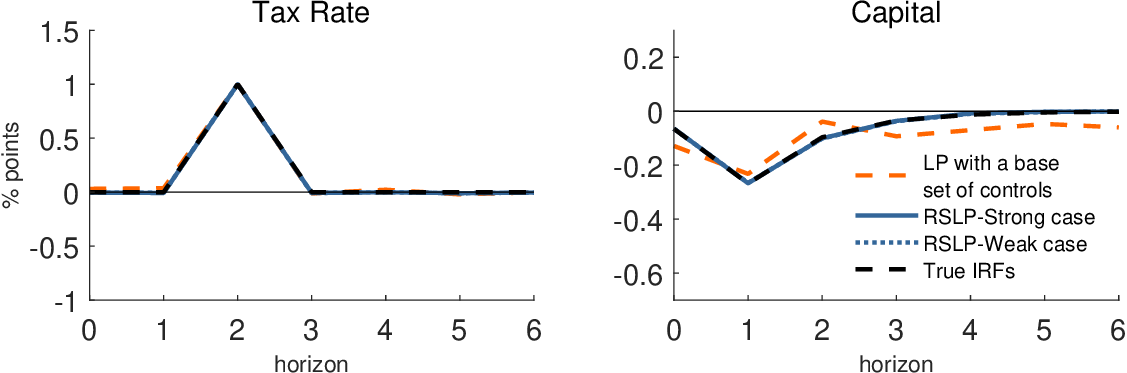}\\
(b) Conditionally exogenous instrument
\includegraphics[width=\textwidth,trim={10 0 10 0}]{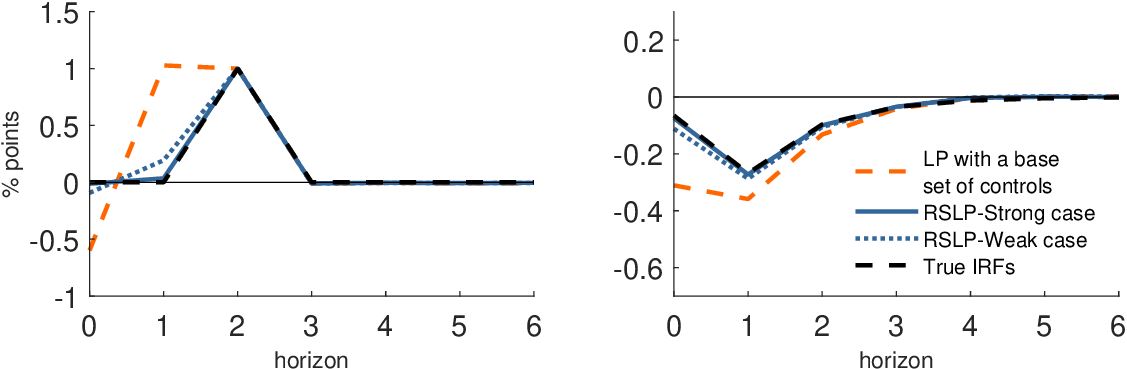}\\
(c) SVAR identification
\includegraphics[width=\textwidth,trim={10 0 10 0}]{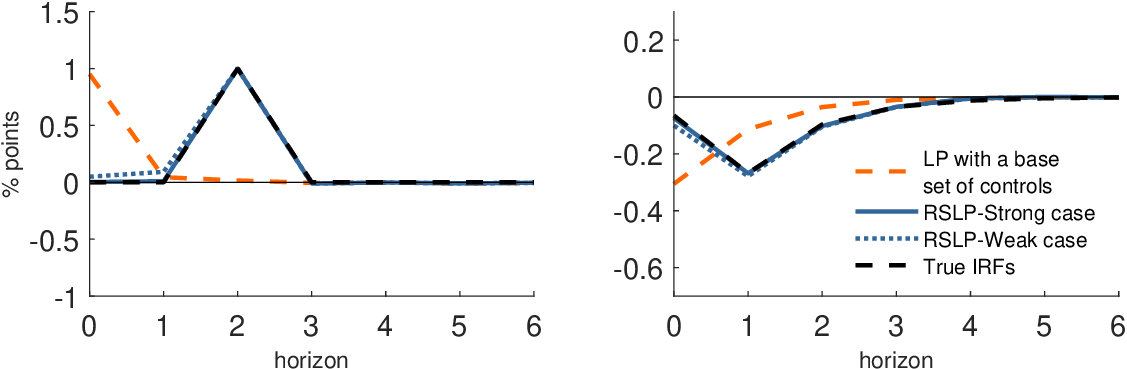}\\
\end{center}
\footnotesize
\emph{Note: This figure shows IRFs estimated by LP with a base set of controls (orange, dot-dashed), RSLP using controls from the strong (blue, solid) and weak case (blue, dotted) in \eqref{eq:inforseriesstrong_weak}, and the true responses (black, dashed). The panels correspond to different identification schemes. The shock is normalized to a 100 basis point increase in the tax rate. The results are the average IRFs across 1000 Monte Carlo replications.} 
\label{fig:fiscallprslporiginal}	
\end{figure}

Figure~\ref{fig:fiscallprslporiginal} presents the IRFs of both capital and tax rate to a tax shock from both the IV and SVAR identification schemes, under both the strong and weak case. We also plot IRFs estimated via LP without the additional controls from the generated informational series, and the true IRFs. The plotted estimated IRFs are taken as the average across the Monte Carlo replications, so deviations from the true IRF represent bias.

Since the tax shock occurs in period 2, the true IRF of the tax rate sees a one off spike in period $t+2$. Capital falls for two periods, then adjusts back in a monotonic fashion. Panel (a) in Figure~\ref{fig:fiscallprslporiginal} shows that with a strictly exogenous instrument the true IRFs are recovered. This result holds even without information on the controls, although the dot-dashed orange line shows some more variation around the true IRF. Now suppose we estimate the IRF without controls in Panel (b) and (c). The dot-dashed orange line shows that we will miss the timing of the tax rate, and as a result, also miss the response of capital. LP with a base set of controls and a conditionally exogenous instrument in Panel (b) is biased given the instrument is invalid due to not being exogenous to the lagged shocks. Panel (c) is an illustration of a fiscal foresight issue, where information on only the tax rate and capital is insufficient to recover the true IRF, which has been previously documented \cite[e.g.][]{leeper2013fiscal,forni2014sufficient}.

Our proposed approach is able to largely recover the true IRF. In particular, RSLP is able to estimate the tax rate spiking in period 2, and recover the dynamics of the response to capital. While our proposed method is close to being unbiased in the strong case, the bias is still minimal under the weak case with either IV or SVAR identification.

\subsection{Comparisons relative to other methods}\label{sec:MCbenchmarks}
This section compares RSLP relative to widely used existing methods for constructing IRFs, and to LPs employing alternative dimension reduction methods. More specifically, we consider FAVARs and factor augmented local projections (FALPs), small and medium Bayesian VARs (BVAR),\footnote{The small BVAR only includes the base set of variables; taxes and capital. The medium BVAR considers 20 variables, with another 18 informational variables in addition to the base set. Extant work suggests that with the standard Minnesota type shrinkage and natural conjugate priors, a 20 variable BVAR often produces similar results relative to large BVARs of over hundred variables \cite[see][]{banbura2008large,Morley_Wong2020}. In our setting, a 20 variable BVAR suffices to make our point.} the Bayesian local projection (BLP) of \cite{ferreira2023bayesian}, and LPs exploiting sparsity with either a LASSO penalty or the variable selection method OCMT proposed by \cite{chudik2018one}. Finally, we consider the LP with a base set of controls (Base). Implementation details of these methods are deferred to Appendix~\ref{A:benchmarks}. 

Table~\ref{table:fiscalMSErelativetoLP} shows the root mean squared error (RMSE) of the IRFs estimated by each method relative to the RMSE of the IRFs estimated by RSLP.\footnote{Note that there are seven horizons ($h=0,\dots,6$) in the IRFs, across which we average to calculate the RMSE: $\sqrt{\frac{1}{7}\sum_{h=0}^{6}\frac{1}{1000}\sum_{i=1}^{1000}(\hat{\beta}_{hi}-\beta_{h,\text{true}})^2}$, where $\hat{\beta}_{hi}$ is the estimated IRF at horizon $h$ in the $i$-th replication and $\beta_{h,\text{true}}$ is the true IRF at horizon $h$.} Values above one favour RSLP. These values are reported for the six experiments with the real business cycle model with fiscal foresight, which we refer to as the baseline experiments. We also compare the methods in experiments with a dynamic factor model (DFM) in the spirit of \cite{stock2016dynamic}. Appendix~\ref{A:dfm} discusses the DFM experiments in detail. Here we summarize broad observations from our baseline experiments, noting that the DFM experiments have similar results unless otherwise stated.

\begin{table}[t!]
\small
	\caption{RMSE relative to RSLP in the baseline Monte Carlo experiments}\label{table:fiscalMSErelativetoLP}
    \begin{tabular}{lrrrrrrrrrrrr}
\toprule \toprule
          & \multicolumn{4}{c}{Strict Instrument} & \multicolumn{4}{c}{Conditional Instrument} & \multicolumn{4}{c}{SVAR} \\
           \cmidrule(lr){2-5}\cmidrule(lr){6-9}\cmidrule(lr){10-13}
          & \multicolumn{2}{c}{Strong} & \multicolumn{2}{c}{Weak} & \multicolumn{2}{c}{Strong} & \multicolumn{2}{c}{Weak} & \multicolumn{2}{c}{Strong} & \multicolumn{2}{c}{Weak} \\
           \cmidrule(lr){2-3}\cmidrule(lr){4-5} \cmidrule(lr){6-7}\cmidrule(lr){8-9}\cmidrule(lr){10-11}\cmidrule(lr){12-13}
          & \multicolumn{1}{l}{Tax} & \multicolumn{1}{l}{Cap } & \multicolumn{1}{l}{Tax} & \multicolumn{1}{l}{Cap } & \multicolumn{1}{l}{Tax} & \multicolumn{1}{l}{Cap } & \multicolumn{1}{l}{Tax} & \multicolumn{1}{l}{Cap } & \multicolumn{1}{l}{Tax} & \multicolumn{1}{l}{Cap } & \multicolumn{1}{l}{Tax} & \multicolumn{1}{l}{Cap } \\
          \midrule
          & \multicolumn{12}{c}{LP methods} \\
          \midrule
FALP  & 0.955 & 0.948 & 1.046 & 1.039 & 0.964 & 0.954 & 1.259 & 1.031 & 0.886 & 1.042 & 1.156 & 1.100 \\
    LASSO & 4.523 & 4.298 & 8.167 & 15.219 & 6.664 & 1.789 & 3.931 & 1.666 & 1.867 & 1.012 & 2.002 & 1.001 \\
    OCMT  & 1.340 & 1.341 & 1.279 & 1.301 & 0.974 & 1.002 & 0.737 & 0.926 & 1.194 & 1.183 & 0.748 & 1.069 \\
    BLP   & 7.093 & 4.380 & 6.125 & 3.761 & 8.280 & 4.284 & 4.603 & 3.824 & 7.164 & 1.683 & 5.121 & 1.633 \\
    Base  & 9.634 & 17.815 & 8.319 & 15.295 & 7.283 & 1.886 & 4.049 & 1.683 & 7.336 & 1.744 & 5.243 & 1.693 \\
    \midrule
          & \multicolumn{12}{c}{VAR methods} \\
          \midrule
    FAVAR & 5.779 & 3.361 & 5.075 & 2.790 & 0.871 & 0.783 & 2.340 & 1.281 & 0.824 & 1.035 & 1.197 & 0.998 \\
    Small  & 7.042 & 4.314 & 6.081 & 3.704 & 8.219 & 4.215 & 4.570 & 3.762 & 7.142 & 1.564 & 5.105 & 1.518 \\
    Medium & 4.964 & 1.557 & 4.385 & 1.241 & 4.996 & 1.303 & 2.535 & 1.210 & 2.377 & 1.284 & 2.639 & 1.497 \\
\bottomrule \bottomrule
    \end{tabular}%
    \footnotesize\\

{\emph{Note: Root mean squared error (RMSE) relative to RSLP. Strict and conditional instruments refer to instruments generated by \eqref{eq:z_exo} and \eqref{eq:z_endo}, respectively. SVAR refers to using an 
SVAR identification scheme. Strong and weak refers to the cases in \eqref{eq:inforseriesstrong_weak}. FALP refers to a factor augmented LP. LASSO and OCMT \cite[see][]{chudik2018one} refer to these procedures applied to estimating the LP. BLP refers to the Bayesian LP by \cite{ferreira2023bayesian}. Base refers to an LP with the base set of controls; taxes and capital. FAVAR refers to a factor augmented VAR. Small and medium refers to respectively a bivariate BVAR of tax and capital and a 20 variable BVAR with tax, capital and 18 informational variables. See text and Appendix \ref{A:benchmarks} for further details.}
}
\end{table}%

\paragraph{RSLP improves upon factor models when the factor structure is weakened}
We focus our initial comparisons relative to factor models given its established tradition in the macroeconomics literature. We first discuss the FAVAR, since this model is shown to be able to recover the IRF in the baseline DGP with SVAR identification \cite[see][]{forni2014sufficient}.\footnote{The FAVAR augments a bivariate VAR system containing taxes and capital with the first two principal components of the informational variables. The IV identification case uses the instrument as per the approach by \cite{mertens2013dynamic} and \cite{gertler2015monetary}. The SVAR identification case identifies the shock by assuming there is no cumulative effect of the technology shock on the tax shock.} 
Table~\ref{table:fiscalMSErelativetoLP} shows that in the strong factor case in the conditionally exogenous instrument or SVAR identification schemes, the FAVAR models have a lower or similar RMSE relative to our RSLP approach.\footnote{Note that we focus the comparison of the FAVAR to the RSLP to these two cases as the FAVAR is invertible. The FAVAR is non-invertible in the strictly exogenous instrument case in our baseline DGP as the controls capture noise in the instrument, explaining its underperformance.} However, the conclusions flip when we consider the weak cases, where RSLP outperforms the FAVAR, at times by substantial margins.

The comparison between FAVAR with RSLP may be driven by differences between a VAR and LP, or between using factors or random subspaces. Comparing RSLP with FALP isolates the latter. Although the margin of improvement decreases, RSLP still outperforms FALP in the weak case. 
This result suggests that the outperformance of the RSLP relative to FAVAR is both driven by the VAR structure and possible issues related to the factors not being able to account for all relevant information. We conclude that with a strong factor structure, such as cross-country data on asset prices and interest rates, factor models do well. RSLP becomes useful with a weaker factor structure, resembling that of a macroeconomic dataset such as the FRED-MD. The DFM experiments also show results in favour of RSLP over FALP. However, since the DGP in those experiments is a closer approximation to a FAVAR, FAVAR is outperforming both.

\paragraph{RSLP does well compared to other local projection methods} 
RSLP is competitive compared to applying other dimension reduction methods to LPs. Together with FALP, RSLP often performs best in terms of RMSE. Hence, depending on the strength of the underlying factor structure in the data, FALP also appears to be a reasonable method. Variable selection with LASSO does not appear to outperform the FALP or RSLP. OCMT performs well with a conditionally exogenous instrument or SVAR identification, but is less accurate  when the instrument is strictly exogenous. Note that, without any guidance from the literature on how to implement variable selection methods within LPs, improvement may be made in selecting the tuning parameters in these methods.

Comparing the base LP with the LPs using dimension reduction in the experiments with a strictly exogenous instrument shows that using a large set of controls often leads to efficiency gains. All the LP methods that use large information sets do similarly or better than an LP with a standard set of controls. Since the controls are not required to identify the shock with a strictly exogenous instrument, any improvements in RMSE are driven by efficiency gains from inclusion of the potentially relevant controls.

In the DFM experiments, both variable selection methods are outperformed by FALP and RSLP. Although BLP is outperformed in the baseline experiments, it performs well in the DFM experiments. The BLP is shrinking towards a BVAR. While a BVAR is a poor approximation in the baseline DGP due to information insufficiency, a BVAR is a reasonable approximation to the DFM DGP, and so correspondingly, the BLP does much better in the latter setting. Therefore, while the BLP is a valuable addition to the empirical toolkit, our exercise also shows that how well the BLP does in practice depends on the underlying BVAR that one chooses to shrink towards.

\paragraph{RSLP may compare favourably relative to VARs, depending on the DGP} 
Relative to VAR methods other than FAVAR, RSLP performs well across the baseline experiments. Within these DGPs, the small BVAR suffers from information insufficiency. The medium BVAR is technically information sufficient, and hence the loss of accuracy relative to RSLP is driven by two features. First, the VARs include equations for all included variables and are therefore vastly overparametrized. Second, to prevent the overparametrization to lead to substantial estimation uncertainty, the VAR methods apply shrinkage to all coefficients. This shrinkage does not discriminate between essential and potentially relevant controls, which may induce a bias. 

Note that the results in Table~\ref{table:fiscalMSErelativetoLP} are based on mean squared errors averaged across horizons. It is well known that the relative accuracy of VARs improves when the horizons increase. Appendix~\ref{A:MCadd} shows that this is also the case in our baseline experiments. For instance, the relative performance of RSLP and FALP stays relatively constant across horizons, while the relative accuracy of FAVAR to RSLP improves with larger horizons. 

While the baseline experiments show that RSLP can be a serious competitor relative to VAR methods, the DFM experiments show that this is not a general result. In the DFM experiments, the BVAR produces accurate IRFs for most variables considered, and also BLP outperforms the LP methods for most variables. The medium BVAR performs better than RSLP across all variables. These results suggest that VARs are a better approximation to the DFM than to the baseline DGP. The DFM DGP generates the data from factors without distinguishing essential variables, making the application of dimension reduction to all variables appropriate. This may benefit VAR models by reducing both the variance due to potential overparametrization and bias due to the shrinkage priors.

\subsection{Take-aways}
We summarize two key take-aways from our Monte Carlo exercise.  First, RSLP is capable of recovering the true IRF, which at least provides \textit{prima facie} evidence that it is a viable approach for applied work. In particular, RSLP is close to being unbiased across a range of different Monte Carlo experiments. Second, relative to a set of alternative benchmarks RSLP at least performs competitively, suggesting it to be an appropriate addition to the standard toolkit for applied macroeconomists.

\section{Empirical applications}\label{sec:empapplications}
We use RSLP to estimate the dynamic responses to technology and monetary policy shocks. Both applications can be traced to a broad empirical literature, where there are suggestions that baseline specifications with minimal controls are insufficient. Hence, these applications are natural settings to understand whether random subspace is a useful method to incorporate information from a large set of controls. 

\subsection{Technology shock application} 
Since \cite{gali1999technology}, a strand of the SVAR literature has investigated the impact of technology shocks on a variety of macroeconomic variables, with a focus on labor market variables \cite[e.g.,][]{francis2005technology,forni2014sufficient,barnichon2010productivity}. Keeping in the spirit of \cite{gali1999technology}, the technology shock is identified as being the only shock that has a long-run impact on labor productivity.\footnote{For settings that we are working with LPs instead of SVARs, we implement the long-run identification restrictions as suggested by \cite{plagborg2021local}. Implementation requires one to nominate a horizon at which all short-run effects of the shock are expected to dissipate. We set this horizon to 3 years. Appendix~\ref{sec:SVARidentification} discusses further implementation details.} 

Mirroring the baseline bivariate VAR in the broader literature, our baseline set of controls includes four lags of both the growth rate of labor productivity and unemployment in ${V}_t$ as we use quarterly data. The set of possible controls in ${G}_t$ includes 127 macroeconomic time series specified at the quarterly frequency, which includes 117 series from the FRED-QD database \citep[see][]{mccracken2020fred}, 6 total factor productivity series from Fernald's website,\footnote{https://www.johnfernald.net/TFP} and 4 consumer confidence indicators from the Michigan Survey. We consider the first lags for these variables, which gives us a set of 127 possible control variables. Our sample is from 1960Q1 to 2019Q4.

\subsection{Monetary policy shock application} 

This application features a baseline set of controls that echoes the external-VAR of \cite{gertler2015monetary}. As the model is monthly, our baseline set of controls includes 12 lags of the log difference of CPI, log difference of industrial production (IP), the excess bond premium by \cite{gilchrist2012credit}, and the 1-year government bond rate in ${V}_t$. 

Our possible control set ${G}_t$ includes 111 FRED-MD series, and we consider the first lag of these variables. The sample spans January 1990 to June 2012 to match up with the time-span of the instrument. We use the high frequency surprises around announcements by \citet{gertler2015monetary} as an instrument to identify the effect of monetary policy shocks.

\subsection{Results}
\begin{figure}[tph]
\begin{center}
\caption{Impulse response functions in the empirical applications}\label{fig:techshocksLPRSLP}
\footnotesize (a) Response to a technology shock
\includegraphics[width=.95\textwidth,trim={50 0 50 0},clip]{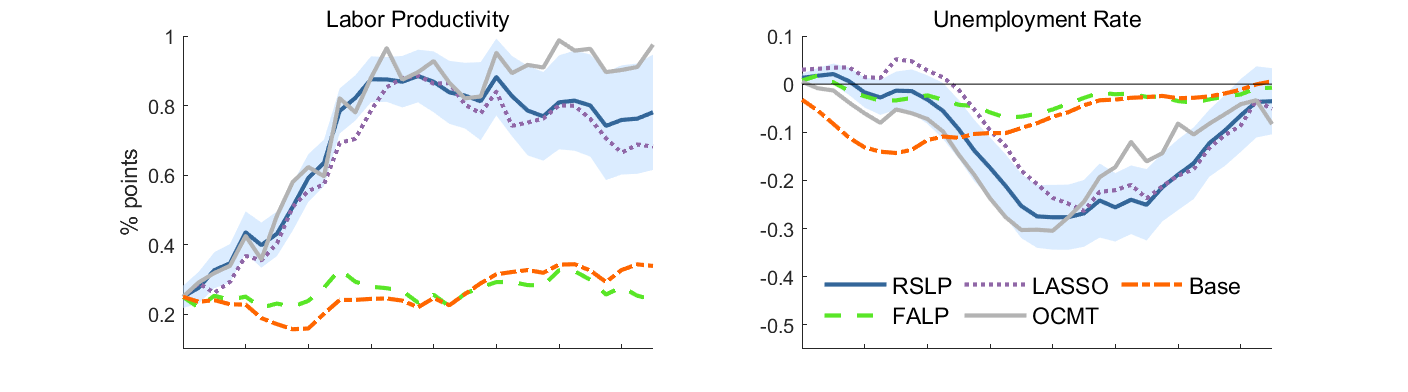}\\ \vspace{-5mm}
\includegraphics[width=.95\textwidth,trim={50 0 50 0},clip]{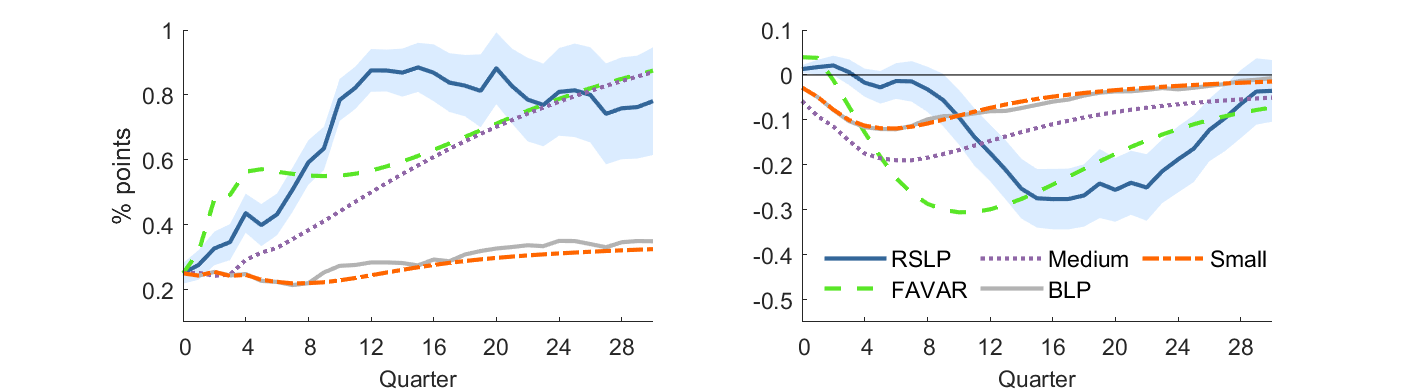}\\
\footnotesize (b) Response to a monetary policy shock 
\includegraphics[width=.95\textwidth,trim={50 0 50 0},clip]{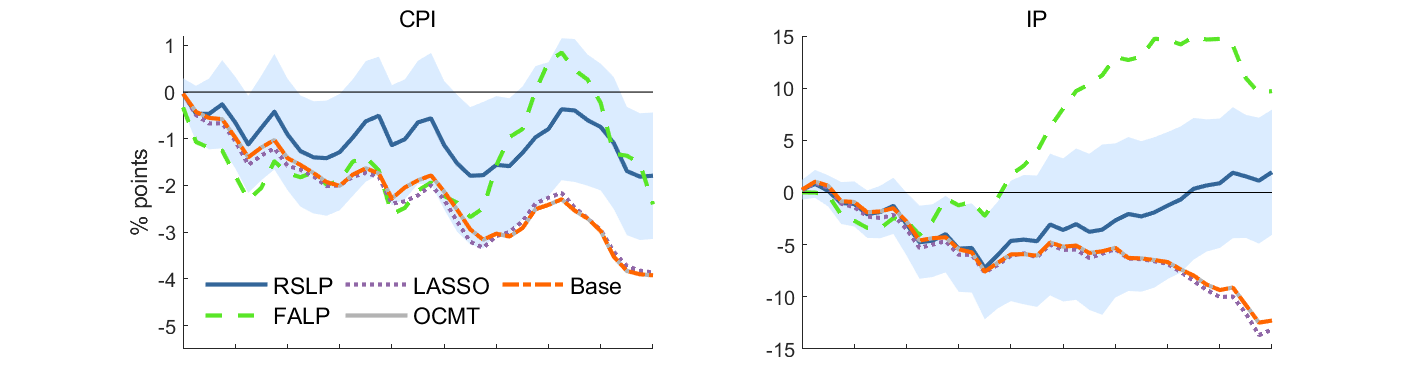}\\ \vspace{-5mm}
\includegraphics[width=.95\textwidth,trim={50 0 50 0},clip]{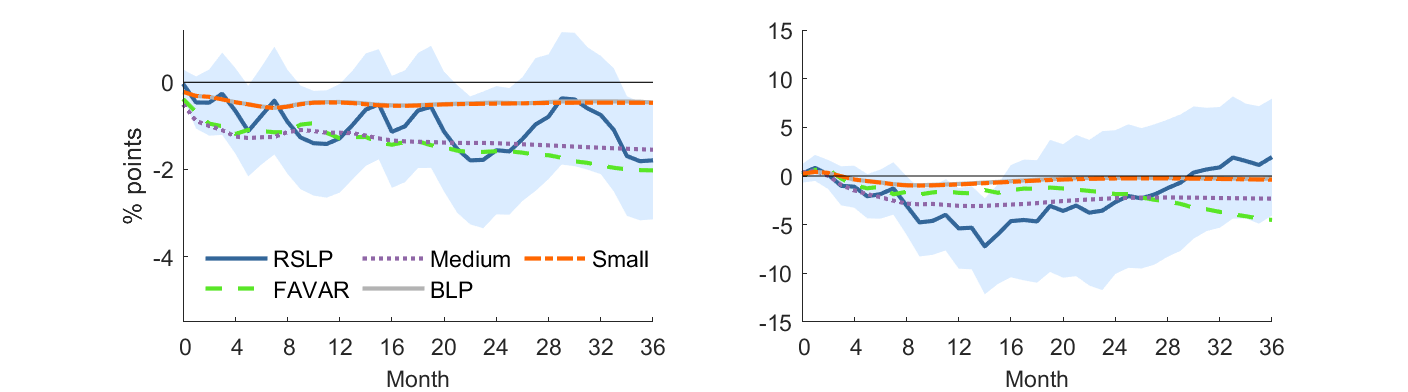}\\
\end{center}
\vspace{-5mm}
\footnotesize
\emph{Note: Panel (a) presents estimated IRFs of labor productivity and unemployment rate to a technology shock which increases labor productivity by 0.25$\%$ on impact. Panel (b) presents estimated IRFs of CPI inflation and industrial production to a monetary policy shock which increases the one year bond rate by 100 basis points. The blue-shaded areas indicate the 90\% error bands of RSLP. See text and note to Table \ref{table:fiscalMSErelativetoLP} for details of models estimated.
} 
\end{figure}

Figure~\ref{fig:techshocksLPRSLP} presents the estimated IRFs from our two applications.\footnote{Note that the IRFs are on the level of labor productivity, CPI and industrial production index. Hence, we follow the usual practice to specify the left-hand side variable in the LP as $y_{t+h}-y_{t-1}$ \cite[see][Section 1.5]{stock2018identification}.} The RSLPs are based on 1000 draws of the selection matrix, a subspace dimension equal to 50,\footnote{Appendix~\ref{A:dimension_monte_carlo} investigates the robustness of the results with respect to the subspace dimension.} and accompanied by an approximate 90\% bootstrap interval constructed as in Appendix~\ref{sec:RSLPsd}. The IRFs are also estimated by the benchmark methods discussed in Section~\ref{sec:MCbenchmarks}. The implementation details of these methods in both applications are deferred to Appendix~\ref{A:benchmark_apps}.

For the technology shock application, we present the response of labor productivity and the unemployment rate to a technology shock which raises labour productivity by $0.25\%$, which is equivalent to a one standard deviation shock in the FAVAR model. Using RSLP, labor productivity increases and the unemployment rate decreases in response to an expansionary technology shock. These results are consistent with the predictions of the real business cycle model and those shown by, for example, \cite{christiano2003happens}.

For the monetary policy shock application, we present the response of CPI and industrial production to a contractionary monetary policy shock which increases the 1-year government bond rate by a 100 basis points. Using RSLP, both CPI and industrial production fall in response to a contractionary monetary policy shock, consistent with what one expects from a standard monetary policy shock.\footnote{The information effect does not seem to produce puzzles here. We nonetheless note that our approach may be well-suited for dealing with informational effects, as variables representing central bank information can be used as additional controls. We experimented with using the 27 variables representing forecasts and forecast revisions constructed by \cite{miranda2018transmission} as part of a larger set of possible controls for our monetary policy application, and obtained almost identical results, which suggests that our large information set already accounted for possible informational effects.}

Comparing relative to other methods, we find that while the direction of the response of the variables is largely consistent across all the different methods, our proposed RSLP also leads to noticeable differences relative to the other approaches. These differences can be qualitatively meaningful, as the estimated responses from the different methods can be outside the bootstrapped 90\% intervals for RSLP. 

More specifically, we find that the relative differences across LP methods vary across the two applications. FALP does not differ much from the base LP in the technology application, but produces IRFs different from all other methods with the monetary policy shock. The IRFs of the LPs with variable selection are similar to the ones from RSLP in the technology application, but do not pick up any information from the controls in the monetary policy shock example.
From the comparison with the VAR methods, we find that FAVAR and the medium BVAR are closer to RSLP than BLP and the small BVAR across both applications, indicating that they pick up some signal from the controls. The BLP can be quite similar to the estimated response from the small BVAR, which highlights the finding from our Monte Carlo experiments that the performance of BLP can be driven by the choice of the BVAR one chooses to shrink towards.

Overall, our empirical exercise suggests that RSLP does lead to reasonable empirical results, especially when compared with other methods and the broader literature. Consistent with our Monte Carlo experiments, we are encouraged that our approach can be a useful addition to the applied macroeconomist's toolkit.

\section{Conclusion}\label{sec:Conclusion}
We show how one can apply a dimension reduction technique traditionally used in machine learning to LPs in order to estimate IRFs with many controls. This random subspace method is simple to implement and it basically contains 3 steps: Step 1: take a random draw; Step 2: do it many times; and Step 3: average. We have shown that our approach is a plausible addition to the toolkit as it can recover the true IRFs in settings encountered in macroeconomics. 

It is worth stressing that while our proposed approach is a plausible empirical strategy, it does not compete to supplant any of the recent innovations in the broader development of LP estimation. To highlight some important developments in the LP literature, \cite{barnichon19smooth} consider smoothing LP, and \cite{ferreira2023bayesian} combine LPs with additional prior information. Both approaches can be applied to RSLPs instead of standard LPs with a base set of controls, so providing a more appropriate starting point for their procedures. Since RSLP averages across estimates from standard LP regressions, each regression can in principle be corrected for finite-sample bias as proposed by \cite{herbst2021bias}, or estimated by generalized least squares as discussed by \citet{lusompa2023local}. 

We also note that while we show that our method can outperform factor models in some empirical plausible settings, we do not argue that one has to choose between either factor or random subspace methods. For instance, there is nothing to stop one from using factors in conjunction with subspace methods. One possibility is the application of subspace methods to a set of estimated factors instead of the original set of controls. Therefore, one should not necessarily view our work as a replacement for existing methods. Instead, we are keen to stress the potential for future work to combine our insights with existing developments in the LP literature in attempts to further improve the properties of these LP estimators.

    \bibliographystyle{jpe}
\bibliography{reference}

\clearpage
\newpage
\appendix

\section*{ONLINE APPENDIX}

\renewcommand{\thefigure}{A\arabic{figure}}
\setcounter{figure}{0}

\renewcommand{\theequation}{A\arabic{equation}}
\setcounter{equation}{0}

\renewcommand{\thetable}{A\arabic{table}}
\setcounter{table}{0}

\section{Selecting and weighting random subspaces}
This appendix discusses three different extensions to the RSLP method discussed in Section~\ref{sec:rslp}:  (1) the random subset selection matrices take predictor categories into account; (2) the subspaces are weighted by Bayesian information criterions (BICs); and (3) the subspace dimension is selected using BIC. 

\subsection{Randomly selecting subspaces using predictor categories}\label{A:categories}

Most macroeconomic applications include datasets in which the variables can be categorized. For instance, in the FRED-MD database, the variables have clear categories, e.g., prices, employment, production, etc. In cases that the researcher has prior knowledge that these categories --or a subset of them-- are essential to the LP, every random subset selection matrix should at least include one predictor from each category. Prior knowledge on which categories are essential may be more common than prior knowledge on which specific predictors are essential. In the latter case the  essential predictors should be included in $V_t$.

Suppose the vector of possibly relevant controls $G_t$ is characterized by $C$ mutually exclusive categories. Without loss of generality, $G_t$ can now be written as $G_t=(G_{1t}',G_{2t}',\dots,G_{Ct}')'$, with $p_{G,c}\times1$ vectors $G_{ct}$ for each category $c=1,\dots,C$. 
To ensure at least one predictor is selected per category by each random subset selection matrix, the matrix $R^{(j)}$ is constructed as follows. First, draw a $k_c \times p_{G,c}$ subset selection matrix $R_c^{(j)}$ for each category $c$. Second, construct the $k \times p_G$ matrix $R^{(j)}=\text{blockdiag}(R_1^{(j)},R_2^{(j)},\dots,R_C^{(j)})$, where $\text{blockdiag}(A,B)$ is a block diagonal matrix with the matrices $A$ and $B$ on its diagonal. The subspace dimension for category $c$ is $1 \leq k_c \leq p_{G,c}$ and $k=\sum_{c=1}^C k_c$ with $C \leq k_c \leq p_{G}$.
We set $k_c$ to the nearest integer such that it reflects the distribution of number of predictors across the categories, and set $k=50$. Alternatively, categories which are a priori considered to be more important can be assigned a relatively large $k_c$. 

\subsection{Weighting random subspaces using the BIC}\label{A:weights}
\citet{elliott2013complete} and \citet{elliott2015complete} have experimented with alternative weighting schemes to the equal weights in \eqref{eq:irfrslp} in a forecasting context. They show that weights proportional to the exponential of the BIC value of a subspace do not result in substantial improvements over equal weighting schemes. This result is consistent with findings in the general forecast combinations literature, usually explained by the additional variance introduced by estimating the weights. Due to the small sample sizes in macroeconomic applications and the relatively large subspace dimension of 50, estimating optimal weights in our setting is therefore also not expected to work well.

Since most identification schemes require the LP to be estimated in two-stages, as is also the case for our empirical applications, the computation of the BIC is not as straightforward as in a predictive regression. Omitting information required for identification in the first stage result in an identification bias. Therefore we estimate the weights based on the BIC from the first stage which does not depend on the horizon $h$. 

Suppose the first stage equation corresponding to draw $R^{(j)}$ is
 \begin{align} 
 {x}_{t}&=\alpha^{(j)}+ \rho^{(j)} z_t + {\Lambda}^{(j)} {V}_t+{\Upsilon}^{(j)}{R}^{(j)}{G}_{t}+\eta_t^{(j)},   
 \end{align}
which is \eqref{eq:rslpiv1} in the main paper. The BIC corresponding to draw $R^{(j)}$ now equals
\begin{align}
    \text{BIC}_j = T \ln\left( \frac{1}{T}\left({x}_{t}-\left(\hat{\alpha}^{(j)}+ \hat{\rho}^{(j)} z_t + \hat{\Lambda}^{(j)} {V}_t+\hat{\Upsilon}^{(j)}{R}^{(j)}{G}_{t}\right)\right)^2 \right) +\ln(T)(2+p_V+k),
\end{align}
where $T$ is the number of observations.
The weighted IRF is constructed as
\begin{align}
    \hat{\beta}_h=\sum_{j=1}^{n_R} \frac{\exp(-\text{BIC}_j)}{\sum_{j=1}^{n_R}\exp(-\text{BIC}_j)} \hat{\beta}_h^{(j)}.
\end{align}
The weight of each subspace is proportional to the exponential of its BIC, which results in larger weights for subspaces with a high likelihood value.

\subsection{Selecting the subspace dimension by BIC}\label{A:selection}
The accuracy of RSLP relies on an appropriate choice of subspace dimension. Papers proposing random subspace methods in forecasting, have selected the subspace dimension recursively based on past predictive performance \citep{elliott2013complete,boot2019forecasting}. Since the IRFs are unobserved, this approach does not work in our setting. 
\citet{elliott2015complete} also select the subspace dimension recursively, but use the BIC instead. In principle, the BIC can be computed for the RSLP, and subsequently be used to select a subspace dimension as well. However, as \citet{elliott2015complete} also note, selection based on the BIC relies on estimates of the residual variance, and hence is not expected to work well when the subspace dimension is large relative to the sample size. Therefore, selection by BIC might not work well in most macroeconomic applications of the RSLP. 

Similar to weighting subspaces with BIC, we construct the BIC corresponding to the first stage to minimise the risk that we omit variables key to the identification of the IRF:
\begin{align}
    \text{BIC}_k = T \ln\left( \frac{1}{T}\left({x}_{t}-\frac{1}{n_R}\sum_{j=1}^{n_R}\left(\hat{\alpha}^{(j)}+ \hat{\rho}^{(j)} z_t + \hat{\Lambda}^{(j)} {V}_t+\hat{\Upsilon}^{(j)}{R}^{(j)}{G}_{t}\right)\right)^2 \right) +\ln(T)(2+p_V+k).
\end{align}
The BIC is considered for a grid $k \in \{0,10,20,30,40,50,60\}$, and the $k$ corresponding to the smallest BIC is selected. Since the first stage usually does not depend on $h$, the selected subspace dimension is the same across all horizons. 

\subsection{Empirical results}
We estimate the IRFs in the empirical applications using the three extensions to the RSLP method, and compare the estimates to the estimated IRFs by RSLP. 

\begin{figure}[h!]
\begin{center}
\caption{Estimated IRFs by RSLP and its extensions in the empirical applications}\label{fig:extension_irf}
(a) Response to a technology shock
\includegraphics[width=\textwidth,trim={00 0 00 0}]{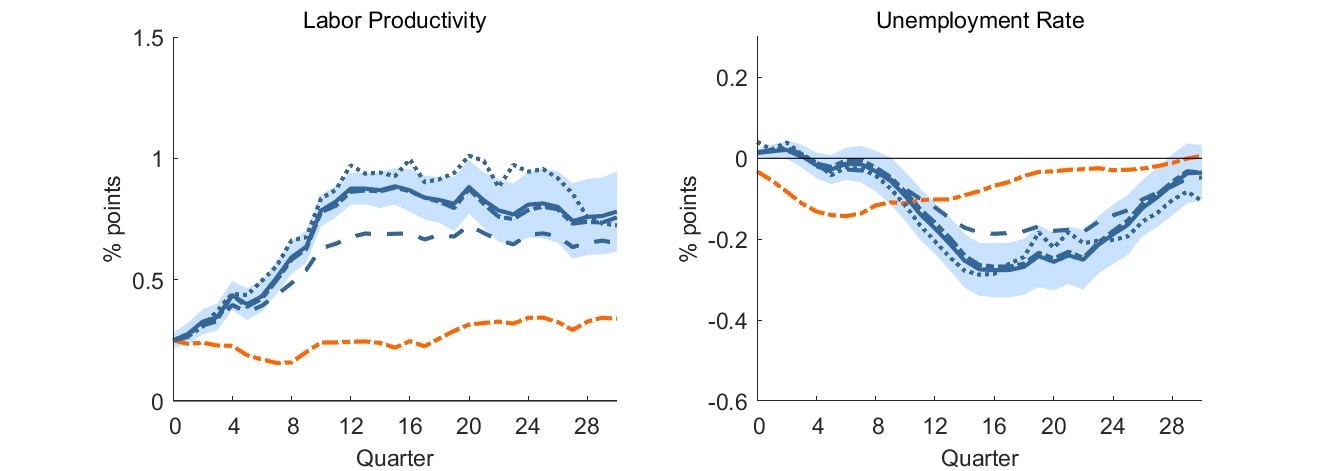}\\
(b) Response to a monetary policy shock
\includegraphics[width=\textwidth,trim={00 0 00 0}]{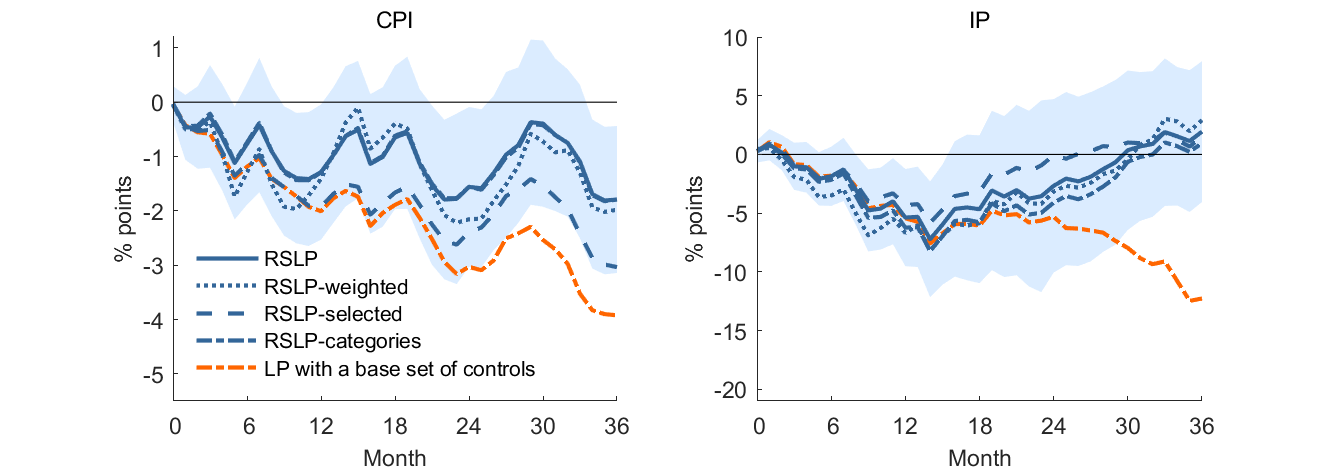}\\
\end{center}
\footnotesize
\emph{Note: This figure shows the estimated LP with only a base set of controls (orange, dot-dashed), RSLP (blue, solid), RSLP with subspaces weighted by BIC (blue, dotted), RSLP with subspace dimension selected by BIC (blue, dashed), and RSLP using predictor categories (blue, dot-dashed). The upper panel presents estimated IRFs of labor productivity and unemployment rate to a technology shock normalized to increase labor productivity by 0.25$\%$ on impact. The lower panel presents estimated IRFs of CPI and IP to a monetary policy shock increasing the one year bond rate by 100 basis points. The blue-shaded areas indicate the 90\% error bands of RSLP.} 
\end{figure}

First, consider the random selection of subspaces using predictor categories. The technology shock application has 127 controls in $G_t$, from which 117 are from FRED-QD, 6 from the total factor productivity series at Fernald’s website, and 4 from the Michigan Survey consumer confidence indicators. These 127 controls can be categorized into 12 categories: NIPA (1), Industrial Production (15), Employment and Unemployment(28), Housing (11), Prices (22), Interest Rates (18), Money and Credit (10), Exchange Rates (4)
Stock Market (5), FRED's Other Series (3), Total Factor Productivity (6), and Michigan's Consumer Index (4), where the number of controls in each category is included in parentheses. The monetary policy shock application has 111 controls from FRED-MD in $G_t$. These controls can be categorized into 8 categories: Output and Income (14), Labor Market (30), Consumption and Orders (10), Orders and Inventories (9), Money and Credit (13), Interest Rates and Exchange Rates (19), Prices (12), and Stock Market (4). Figure~\ref{fig:extension_irf} shows that the estimated IRFs from randomly selecting subspaces using these categories and without the categories are almost identical. 

Second, Figure~\ref{fig:extension_irf} shows the estimated IRFs from RSLP with equal weights and a fixed subspace dimension of 50, together with RSLP with weights based on the BIC and RSLP with a subspace dimension selected by BIC. The estimated IRFs with equal weights and BIC weights are very similar in our empirical applications. The estimated IRFs from RSLP with a subspace dimension selected by BIC is either close to RSLP with a fixed subspace dimension of 50 or in between this RSLP and the LP with a base set of controls, due to the BIC often selecting small subspace dimensions.

\section{Subspace dimension}\label{A:dimension_monte_carlo}
\subsection{Subspace dimension in the Monte Carlo experiments}
\begin{figure} [h!]
\caption{Subspace dimension RSLP in the Monte Carlo experiments}\vspace{-5mm}
\begin{center}
 (a) Strictly exogenous instrument
\includegraphics[width=.95\textwidth,trim={10 0 10 0}]{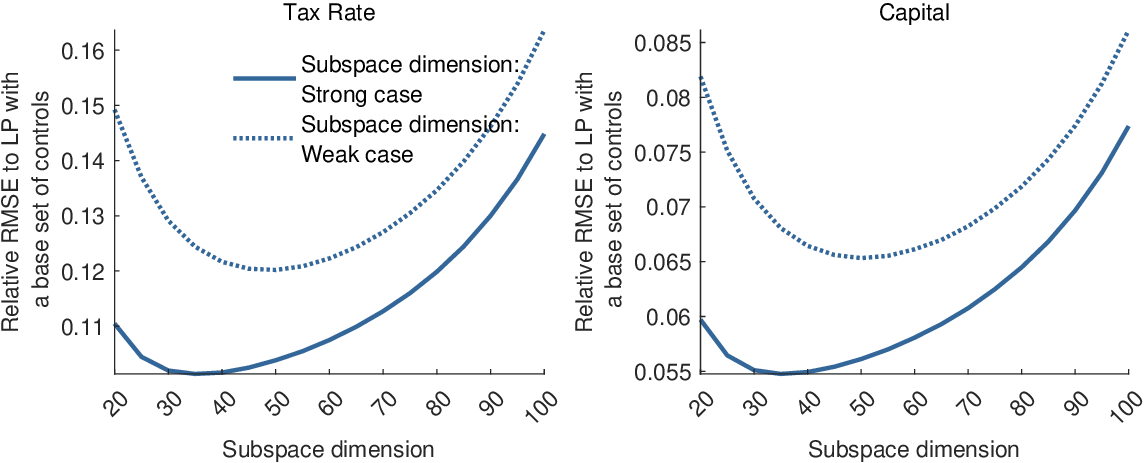}\\
(b) Conditionally exogenous instrument
\includegraphics[width=.95\textwidth,trim={10 0 10 0}]{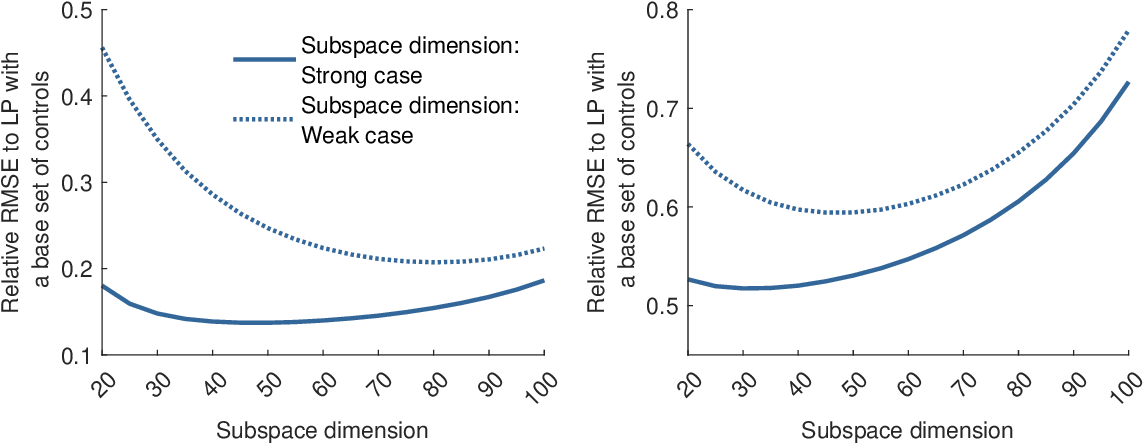}\\
(c) SVAR identification
\includegraphics[width=.95\textwidth,trim={10 0 10 0}]{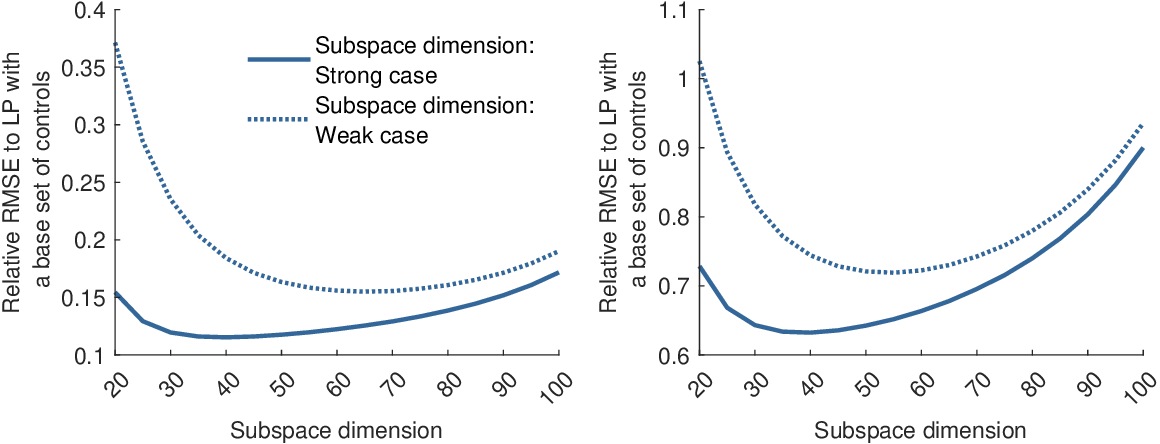}\\
\end{center}
\footnotesize
\emph{Note: This figure shows the root mean squared error of RSLP relative to LP with a base set of controls for various subspace dimensions, in the strong  (blue, solid) and the weak case (blue, dotted) in \eqref{eq:inforseriesstrong_weak}, respectively. The panels correspond to different identification schemes.} 
\label{fig:fiscalrslpdim}
\end{figure}

We explore the size of the subspace dimension in our Monte Carlo experiments. For both variables and the different identification schemes, we calculate the root mean squared error (RMSE) of the RSLP across horizons 0-6 for each subspace dimension. Figure~\ref{fig:fiscalrslpdim} reports these RMSEs relative to the RMSE of the LP with the base set of controls. 

While we do not find a unique subspace dimension that minimizes the RMSEs in each setting under consideration, we highlight two findings from this exercise. First, varying the subspace dimension reveals a ``U-shaped'' pattern, suggesting that when there are a few controls in each subspace regression, additional controls improve the estimates. This is, however, a trade-off where for some subspace dimensions the variance in the estimates exceeds the reduction in bias. Second, we find that the minimum is often achieved with a subspace dimension within the 40-60 range, and the subspace dimension that minimizes the RMSEs is smaller in the strong cases. Intuitively, one needs fewer variables in the subspace regressions to filter out the relevant signal with stronger signals from the informational variables.

In general, these findings are consistent across horizons as well, except for some smaller horizons for tax rate where larger subspace dimensions are optimal. Since the optimal subspace dimension is not known a priori, and Appendix~\ref{A:selection} discusses that selecting this dimension does not necessarily improves the results, it seems reasonable to set $k=50$.

Although this is a simulated data example, we stress that the Monte Carlo experiments are useful in designing empirical strategies on how to set the subspace dimension for applied work. First, the suggestion from the Monte Carlo is that for a dataset with a factor structure like the FRED-MD, a subspace dimension in the range of 40 to 60 is probably a good starting point. Moreover, \cite{boot2020subspace} also find an optimal subspace dimension of 40-60 with the FRED-MD, albeit in a forecasting context. Second, given datasets are different, exploring some form of robustness of the subspace dimension is probably warranted for applied work. Hence, we proceed with an empirical strategy that uses a subspace dimension in the 40-60 range, and subsequently check for the robustness of the results.

\clearpage
\subsection{Subspace dimension in the empirical applications}
\begin{figure} [h!]
\begin{center}
	\caption{Subspace dimension RSLP in the empirical applications} \label{fig:techshockrobust}
 (a) Response to a technology shock
	\includegraphics[width=.95\textwidth,trim={50 0 50 0},clip]{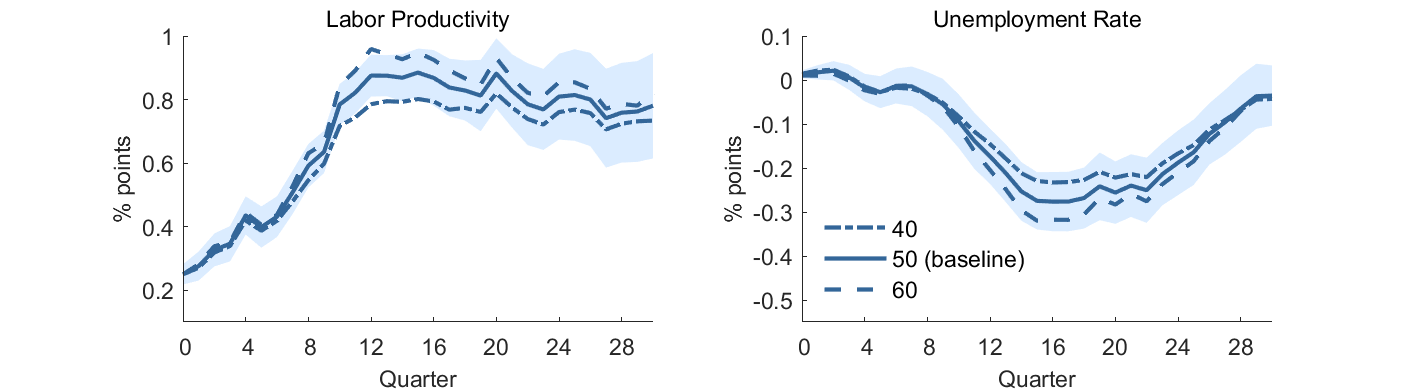}\\
 (b) Response to a monetary policy shock
 \includegraphics[width=.95\textwidth,trim={50 0 50 0},clip]{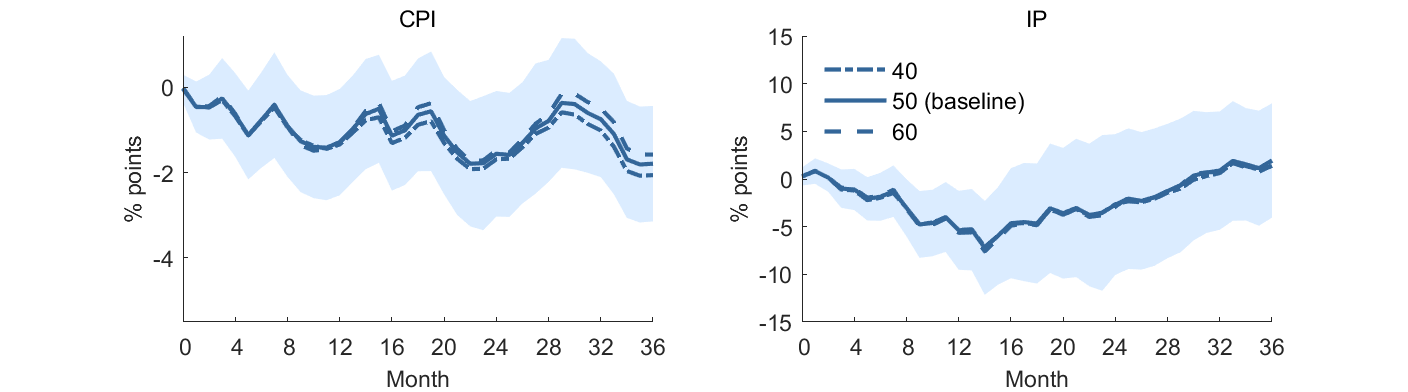}\\
 \end{center}
\footnotesize
\emph{Note: The upper panel presents the estimated IRFs of labor productivity and unemployment rate to a technology shock normalized to increase labor productivity by 0.25$\%$ on impact. The lower panel presents estimated IRFs of CPI and IP to a monetary policy shock increasing the one year bond rate by 100 basis points. The blue-shaded areas indicate the 90\% error bands of RSLP with a subspace dimension of 50.}
\end{figure}

In the two empirical applications, we set the subspace dimension equal to 50 as follows from the discussion above. However, it remains an open question whether this choice of subspace dimension is appropriate in our applications. To investigate this choice, Figure~\ref{fig:techshockrobust} presents the estimated IRFs by RSLP for our applications, but now with varying subspace dimensions. The blue solid-line corresponds to a subspace dimension of 50, which is displayed alongside the estimates corresponding to the subspace dimensions of 40 and 60. We find that these estimated IRFs are very similar to that considered in our baseline. Moreover, they are within the 90\% error bands of the baseline choice of 50. This indicates that our choice of subspace dimension is appropriate.

\clearpage

 \section{Standard deviation of impulse response functions}\label{sec:RSLPsd}

\subsection{Bootstrapping}

To construct error bands via bootstrapping, we first recognize that the estimated IRF $\hat{\beta}_h$ is the average of estimated IRFs $\hat{\beta}_h^{(j)}$ from $n_R$ different regressions. It follows that one can construct the standard error of $\hat{\beta}_h$ as long as one has the variances and covariances associated with each of the $\hat{\beta}_h^{(j)}$'s.

One natural approach is to construct the covariances between the $\hat{\beta}_h^{(j)}$'s via bootstrapping. When designing the bootstrap, one should allow for possible serial correlation up to an order of $h$ in the residuals, as commonly done when one estimates LPs, and also allow for possible correlation across the $\hat{\beta}_h^{(j)}$'s. We use a moving block bootstrap of block size $h$ in order to allow for serial correlation up to an order of $h$. To allow for cross correlation, we draw the same corresponding block across the $n_R$ regressions to preserve any possible cross correlation.

From a given horizon, $h \in \lbrace 0, 1, 2, \hdots, H \rbrace$, with a sample of $T$ observations to estimate the IRF, we first define $T-h$ moving blocks of residuals $\hat{\xi}_{t+h}^{(j)}$, where $j$ denotes the index of the $j^{\text{th}}$ draw of the random subspace algorithm. We sample with replacement $(T/h) + 1$ blocks from the $T-h$ moving blocks. We only draw this once for each bootstrap replication (as opposed to $n_R$ times), then use the same corresponding blocks across all the $n_R$ regressions for the same bootstrap replication.
Using these blocks, we obtain bootstrapped samples for each of the $\hat{\beta}_h^{j}$. These bootstrapped samples are used to form an $n_R \times n_R$ covariance matrix for all $\hat{\beta}_h^{j}$'s, and use these quantities to estimate the standard error of $\hat{\beta}_h$. Using the critical values of a standard normal distribution, we  construct an approximate 90\% interval.

\subsection{Analytical Expression}
Recall that our impulse response estimates are constructed as 
\begin{align}
	\hat{\beta}_{h}=\frac{1}{n_R}\sum_{j=1}^{n_R}\hat{\beta}_{h}^{(j)}.
\end{align}
\cite{buckland1997model} derive an expression for the variance of $\hat{\beta}_{h}$ under two assumptions. First, assume that the expectation across all possible draws $R^{(j)}$ of the misspecification bias in estimating $\beta_{h}^{(j)}$ in the model corresponding to $R^{(j)}$ is zero. That is, $\text{E}[\beta_{h}^{(j)}]=\beta_{h}$. Second, assume that $\beta_{h}^{(j)}$ and $\beta_{h}^{(l)}$ are perfectly correlated, for all $l \neq j$. Although this is a strong assumption, each $\beta_{h}^{(j)}$ is estimated on the same underlying dataset and these correlations are therefore indeed expected to be high.

It now follows that
\begin{align}
    \text{SD}[\hat{\beta}_{h}] = \frac{1}{n_R}\sum_{j=1}^{n_R}\sqrt{\text{var}(\hat{\beta}_{h}^{(j)}|\text{model $(j)$ is correct})+({\beta}_{h}^{(j)}-{\bar{\beta}}_{h})^2},
\end{align}
which may be estimated by replacing $({\beta}_{h}^{(j)}-\bar{{\beta}}_{h})$ by $(\hat{\beta}_{h}^{(j)}-\hat{{\beta}}_{h})$, and $\text{var}(\hat{\beta}_{h}^{(j)}|\text{model $(j)$ is correct})$ by the squared Newey-West standard error for the ordinary least squares estimate $\hat{\beta}_{h}^{(j)}$ in model $(j)$ to account for the serial correlation in the error terms \cite[see][]{jorda2005estimation}. As discussed in the text, it is likely that the assumption of  perfect correlation is too strong in practice, and so the bootstrap may be preferred.

\section{Implementing SVAR identification in RSLP}\label{sec:SVARidentification}
\subsection{Monte Carlo experiments}

Equation \eqref{eq:fiscalforesightDGP} shows that all the variation in the tax rate is from the tax shock (and not the technology shock). Hence, the tax shock can be identified by assuming that the cumulative impact of the tax rate on a technology shock is zero. In practice, this approach corresponds to how the long-run restrictions such as by \cite{blanchard1988dynamic} are implemented since they are based on restrictions on the cumulative effect of shocks.

However, there is an important difference between the SVAR identification we implement here and how long-run restrictions are usually implemented. In long-run restrictions, one has an I(1) variable in which all the variation is driven by the permanent shock. Since the I(1) variable enters the VAR or the LP in differenced form, this implies a restriction where the cumulative impact on the \textit{changes} from the transitory shocks to the I(1) variable is equivalent to zero, and thus has zero long-run impact on the level. In contrast, there is no long-run in our setting since neither the tax rate nor capital is I(1), but we can still identify the technology shock since the long-run cumulative impact of this shock on the tax rate (whether differenced or in levels) is zero. 

The estimation of the cumulative identification in an LP setting is shown by \cite{plagborg2021local}. This method constructs a new tax variable by accumulating our tax rate variable in Equation \eqref{eq:fiscalforesightDGP}: 
\begin{align}
	\hat{\tau}^{\text{accum}}_{1}=\hat{\tau}_{1}, \quad \quad 
	\hat{\tau}^{\text{accum}}_{t}=\hat{\tau}_{t}+\hat{\tau}^{\text{accum}}_{t-1}, \quad \text{for}\; t=2,3,...T. \label{eq:taxlevel}
\end{align} 
Define $\upsilon_t = \hat{\tau}^{\text{accum}}_{t+2}-\hat{\tau}^{\text{accum}}_{t-1}$. 
The IRFs can now be estimated by two-stage least squares in the first and second stage regressions 
 \begin{align} 
 \upsilon_t&=\alpha^{(j)}+{\Lambda}^{(j)} {V}^{1}_t+{\Upsilon}^{(j)}{R}^{(j)}{G}^{1}_{t}+\eta_t^{(j)},   \label{eq:rslpsvar1}\\
 y_{t+h}&=\mu_h^{(j)}+ \beta_{h}^{(j)} \hat{\upsilon}_{t}^{(j)}+{\Theta}_{h}^{(j)}{V}^{2}_t+{\Psi}_{h}^{(j)}{R}^{(j)}{G}^{2}_{t}+\xi_{t+h}^{(j)},  \label{eq:rslpsvar2}
 \end{align}
where ${V}^{1}_t$ contains the contemporaneous values of the tax rate and capital, ${G}^{1}_{t}$ contains contemporaneous values of the informational series, ${V}^{2}_t$ consists of the two lags of tax rate and capital, and ${G}^{2}_{t}$ consists of the first lag of the informational series. 
The first stage uncovers the linear combinations of the data that explain the two-period ahead movement of the tax rate. Therefore the fitted values of the first stage recover the tax shock. Once the tax shock is recovered, one can estimate the impulse responses from the second stage.

\subsection{Empirical application with a technology shock} \label{sec:empirical_LR}
The procedure outlined above also applies to our empirical application to a technology shock, with minor adjustments. First, $\upsilon_t$ now equals the long-run movement of labor productivity in levels: $\upsilon_t = \text{labor}^{\text{level}}_{t+12}-\text{labor}^{\text{level}}_{t-1}$. Second, the controls in both stages are now as follows: The first stage controls in ${V}^{1}_t$ are the first-differences of contemporaneous values of labor productivity and the unemployment rate, and ${G}^{1}_{t}$ contains contemporaneous values of 127 FRED-QD series. The second stage controls in ${V}^{2}_t$ are the first four lags of the first-differences of labor productivity and the unemployment rate, and ${G}^{2}_{t}$ consists of the first lag of the 127 FRED-QD series. 

\section{FRED data structure}\label{sec:datastructure}
\begin{figure}[h!]
\begin{center}
	\caption{Factor structure} \label{fig:factorstructure}
	\includegraphics[width=\textwidth,trim={10 0 10 0}]{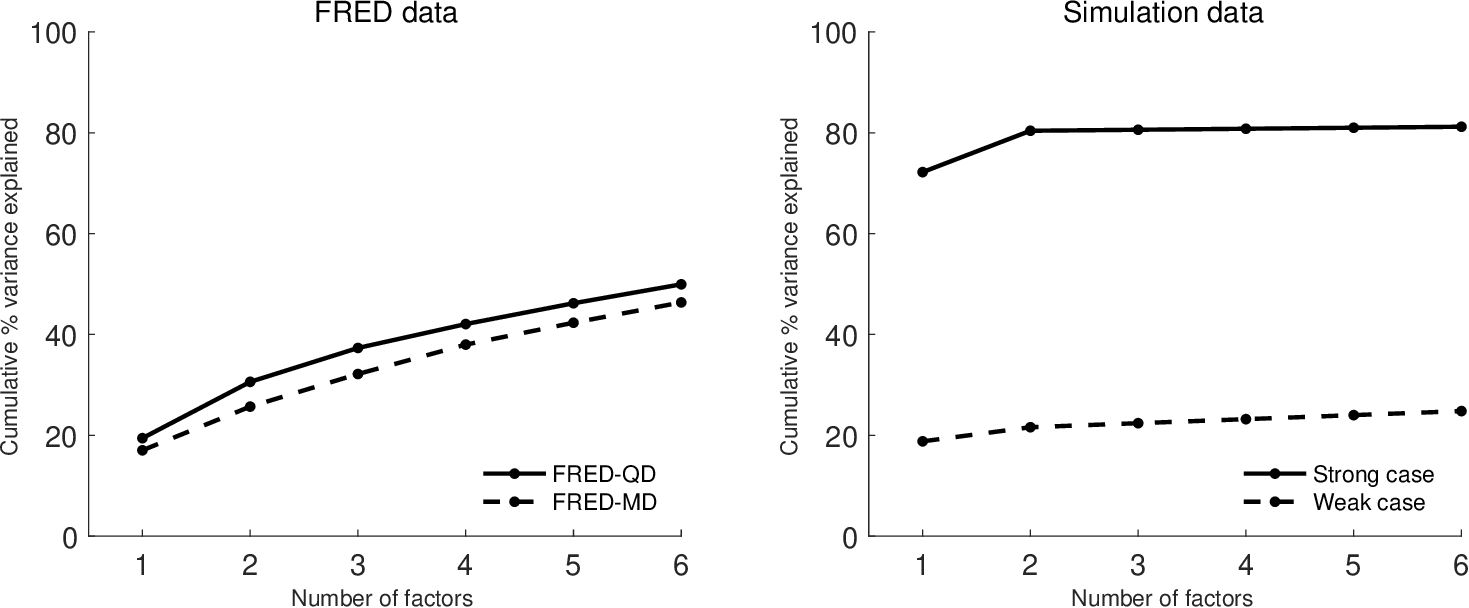}\\
 \end{center}
\footnotesize
\emph{Note: This figure shows the factor structure of FRED-QD, FRED-MD, and our simulated informational series in \eqref{eq:inforseriesDGP} in the strong and weak case.}
\end{figure}

Figure~\ref{fig:factorstructure} shows the factor structure of the FRED-QD and FRED-MD datasets, together with the factor structure of the simulated informational series in \eqref{eq:inforseriesDGP} in Section~\ref{sec:MCsimullongrunexample}. More precisely, we present the cumulative variance being accounted for as we sequentially add an additional principal component. This is a standard metric for understanding the strength of the factor structure. 
The figure shows that the first two factors of FRED-QD and FRED-MD account for 25$\%$ and 20$\%$ of the variation, respectively. Similarly, our simulation data under the weak case also requires two factors to account for 20$\%$ of the variation in the controls. 

Note that in Figure \ref{fig:factorstructure}, we calculate the factor structure of the simulated data in population. Let $n_{y^*}=100$ be the number of informational series, which have mean zero and unit variance. Given the setting in Section \ref{sec:DGP}, the expected number of informational series containing the tax shock equals 10, and the expected number of informational series containing the capital shock equals 90. The population covariance matrix of the informational series is therefore equal to $\Sigma_{y^*}= H+\sigma^2I_{n_{y^*}}$, where $H$ is a $100\times 100$ blockmatrix with a $10\times 10$ and $90 \times 90$ blockmatrix with all elements equal to 1 on its diagonal, $\sigma^2=0.25$ in the strong case and $\sigma^2=4$ in the weak case, and $I_{n_{y^*}}$ is a $100\times 100$ identity matrix. 

Then the factor structure of the dataset can be calculated by solving the eigenvalue-eigenvector decomposition of the correlation matrix $\rho_{y^*}$ of the covariance matrix $\Sigma_{y^*}$: $\rho_{y^*}\nu= \Lambda\nu $,
where $\nu$ is a matrix whose columns are the eigenvectors and $\Lambda$ is a diagonal matrix with the eigenvalues $\lambda_j$ on its diagonal. The cumulative variance explained by the first $i$ principal components can be calculated as $\frac{1}{100}\sum_{j=1}^{i}\lambda_j$, where $\lambda_j$ is the $j^{th}$ largest eigenvalue of $\lambda$.

\section{Benchmark methods in the Monte Carlo experiments}\label{A:benchmarks}

All LP methods estimate \eqref{eq:lpwolf1} with two lags of tax rate and capital in ${V}_t$, together with two lags of the instrument when using IV identification. FALP defines $G_t$ as the first lag of the two principal components that explain the most variation in $\{y^*_s\}_{s=1}^t$.  

The identification schemes in the experiments require the LP to be estimated in two-stages. Since omitting information required for identification in the first stage results in an identification bias, we apply variable selection methods to the first stage. We use both the LASSO as implemented by \citet{glmnet} and OCMT by \citet{chudik2018one} to find a subset of $\{y^*_s\}_{s=1}^t$ to be included in $G_t$. Both the first and second stage regressions are subsequently estimated with the selected variables in $G_t$. 

Conditional on the instrument and $V_t$, LASSO includes $y^*_t$ in $G_t$ in the first stage and adds a LASSO penalty to the corresponding coefficients. All variables with a nonzero coefficient estimate are selected. The penalty parameter is selected from the grid $\exp(-j/2)$ with $j=0,1,\dots,20$, by minimizing the one-observation-ahead out-of-sample errors of the final hundred observations. To construct each error, the model is estimated using an expanding estimation window. OCMT takes $y^*_t$ as the set of potential signals for the first stage, and includes the instrument and $V_t$ as preselected variables. We take the default settings $p=0.05$, $c=1$, $\delta=1$, $\delta^*=2$. 

The VAR methods include tax rate and capital, the instrument is accounted for by the external-IV method \cite[see][]{mertens2013dynamic,gertler2015monetary}, and two lags are considered. The BVAR is implemented according to \citet{giannone2015prior} and the BLP as proposed in \citet{ferreira2023bayesian}. The BVAR uses a natural conjugate prior with a standard Minnesota type structure, with each VAR equation shrunk towards a random walk \cite[see][]{banbura2008large}. The degree of shrinkage is then specified as a single hyperparameter, where the \citet{giannone2015prior} approach calibrates this  hyperparameter by maximizing the marginal likelihood of the model. The BLP approach takes the posterior mode of the BVAR coefficients estimated using the \citet{giannone2015prior} approach as given, and iterates on these coefficients in order to specify a prior on the coefficients of the LP regression using a natural conjugate prior. The degree of shrinkage towards the BVAR coefficients is then estimated by maximizing the marginal likelihood. We shrink the BLP towards the small BVAR in our example as the small BVAR would be considered a ``standard'' specification in each application. We consider two BVARs. The small BVAR includes only the essential variables (i.e. tax rate and capital), whereas the medium BVAR includes 20 variables, by adding 18 information series from $y^*_s$ in addition to the 2 essential variables. The FAVAR includes the two principal components that explain the most variation in $\{y^*_s\}_{s=1}^t$ in addition to the essential variables and is estimated as in \citet{bernanke2005measuring}. 

\section{Dynamic factor model experiments}\label{A:dfm}
This appendix conducts additional Monte Carlo experiments based on a the dynamic factor model (DFM) discussed by \cite{stock2016dynamic}. This model assumes the following process for the $n_X \times 1$ vector of macroeconomic variables $X_t$:
\begin{align}
X_t=\Lambda f_{t}+v_t,\quad 
v_{i,t}=\Delta_i(L)v_{i,t-1}+\Xi_i\xi_{i,t}, \quad  \xi_{i,t}\sim iid \; N(0,1), \label{eq:idioeq}
\end{align}
where the $n_f \times 1$ vector of latent factors follows a stationary VAR model
\begin{align}
f_t=\Phi(L)f_{t-1}+ \vartheta_t, \quad \vartheta_t \sim N(0,\Sigma_\vartheta). \label{eq:factoreq} 
\end{align}

Following \cite{stock2012disentangling}, we set $n_f=6$ and the number of lags in both \eqref{eq:idioeq} and \eqref{eq:factoreq} equal to 4. The parameters in $\Lambda$, $\Delta_i(L)$, $\Xi_i$, $\Phi(L)$, and $\Sigma_\vartheta$ are set to the values estimated on the \cite{stock2016dynamic} dataset consisting of 222 quarterly observations from 1959Q3 to 2014Q4 on $n_X=207$ macroeconomic series.\footnote{We follow \cite{stock2012disentangling,stock2016dynamic} with all the data handling steps and transformations, including the adjustments of outliers, the treatment of missing observations, and detrending using a biweight filter with a bandwidth of 100 quarters.} We empirically calibrate the monetary policy shock, $\epsilon_t^{\text{MP}}$ by estimating a regression of the monetary policy shock instrument by \cite{romer2004new} on each of the empirically estimated residuals $\hat{\vartheta_t}$.

\subsection{Information sets and identification schemes}
The objective is to estimate the IRFs to a monetary policy shock. We consider two experiments. First, the econometrician observes a strictly exogenous instrument. Second, the econometrician observes a conditionally exogenous instrument. In both experiments, we also assume that 105 of the 207 macroeconomic series in $X_t$ are observed, mimicking the size of the information set when working with datasets like FRED-QD or FRED-MD. These 105 series are selected so that all thirteen variable groups in the dataset are represented.

\paragraph{Strictly exogenous instrument} 
First, the instrument $z_t$ for the monetary policy shock $\epsilon^{\text{MP}}_{t}$ is generated as 
\begin{align}\label{eq:z1_dfm}
z_t=\sqrt{0.5}\epsilon^{\text{MP}}_{t}+\nu_t, \quad  \nu_t\sim N(0,1).
\end{align}
The instrument identifies the IRF without controls. However, being able to account for variation in the variable of interest due to movements in the other observed variables from $X_t$ can lead to efficiency gains. Note that this is a slightly different setting compared to the strictly exogenous instrument experiment in Section~\ref{sec:MCsimullongrunexample}. In that experiment, the controls address the variation in the instrument $z_t$ in the first stage, where the controls in the experiment here target variation in the variable of interest $y_t$ in the second stage.

\paragraph{Conditionally exogenous instrument}
Second, the instrument $z_t$ for the monetary policy shock $\epsilon^{\text{MP}}_{t}$ is generated as 
\begin{align}\label{eq:z2_dfm}
z_t=\epsilon^{\text{MP}}_{t}+\epsilon^{\text{MP}}_{t-1}+\eta^{\text{Tax}}_{t-1}+0.9\eta^{\text{Tech}}_{t-1}+\nu_t, \quad  \nu_t\sim N(0,1),
\end{align}
where $\eta^{\text{Tax}}_{t-1}$ and $\eta^{\text{Tech}}_{t-1}$ are respectively tax and technology components of the reduced form forecast errors, $\vartheta_{t-1}$. The crucial point here is because these components are a function of lagged shocks, the instrument is now only conditionally exogenous since it fails lead-lag exogeneity. Using the instrument without controlling for the information in $\eta^{\text{Tax}}_{t-1}$ and $\eta^{\text{Tech}}_{t-1}$ will thus render the instrument invalid. Since the observed variables from $X_t$ are a function of the lagged shocks, they may help recover the IRFs by making the instrument valid conditionally, as long as they control for the information contained in the other shocks in $\eta^{\text{Tax}}_{t-1}$ and $\eta^{\text{Tech}}_{t-1}$.

We empirically calibrate $\eta^{\text{Tax}}_{t}$ and $\eta^{\text{Tech}}_{t}$ using the tax instrument by \cite{romer2010macroeconomic} and the technology instrument by \cite{fernald2014}, respectively. We first obtain the coefficients, $h^{\text{Tax}}$, from a regression of the tax instrument by \cite{romer2010macroeconomic} on the residuals of the monetary policy shock regression which we used to obtain $\epsilon_t^{\text{MP}}$, and subsequently get the coefficients $h^{\text{Tech}}$ by estimating a regression of the technology instrument by \cite{fernald2014} on the residuals from the tax instrument regression. We then form $\eta^{\text{Tax}}_{t}$ and $\eta^{\text{Tech}}_{t}$ as linear combinations of the forecast errors where each of the forecast errors loads on each component. We construct the unnormalized loadings by taking the inverse of the coefficients $h^{\text{Tax}}$ and $h^{\text{Tech}}$. We then normalize by ensuring that the loadings of each component sum to unity. Note that $\eta^{\text{Tax}}_{t}$ and $\eta^{\text{Tech}}_{t}$ are linear combinations of the forecast errors which are functions of other shocks in the model. Hence the instrument fails lead-lag exogeneity without any controls, whether $\eta^{\text{Tax}}_{t}$ and $\eta^{\text{Tech}}_{t}$ are viewed as structural shocks or not.\footnote{The only structural shock which we interpret in the model is the monetary policy shock $\epsilon^{\text{MP}}_{t}$, and deliberately labelled $\eta^{\text{Tax}}_{t}$ and $\eta^{\text{Tech}}_{t}$ as components to make the distinction clear. This simplifies the analysis, as the samples of the different instruments are not aligned and instruments are also not orthogonal in practice \cite[see][for documentation of and discussion of the latter]{stock2012disentangling}.}

\subsection{Methods}
Each experiment simulates 1000 artificial datasets of 200 observations according to \eqref{eq:idioeq} and \eqref{eq:factoreq}, from which only 105 variables are retained. Each artificial dataset is accompanied by an instrument generated from either \eqref{eq:z1_dfm} or \eqref{eq:z2_dfm}.  Since the Monte Carlo experiments are parametrized by empirically estimating the model on data, the artificial data generated from the Monte Carlo thus represent particular variables of interest as they will mimic time series properties of these variables. We therefore refer to these artificial data according to the variables that they represent in the dataset which we used to parameterize the DFM.

LP with a base set of controls includes four lags of the Federal funds rate, industrial production, consumer price index, unemployment rate, and excess bond premium in $V_t$, together with four lags of the instrument. RSLP uses the same specification for $V_t$, and includes the first lag of the remaining 100 observed macroeconomic series in $G_t$. 

The benchmark methods are implemented as outlined in Appendix~\ref{A:benchmarks}, where the macroeconomic time series included in $G_t$ for RSLP are considered as information series $\{y^*_s\}_{s=1}^t$. We include six principal components in FALP and FAVAR, which is the correct number of factors in the data generating process. The small BVAR includes the same variables as in $V_t$ and considers four lags. The medium BVAR includes 20 variables, and hence we select in addition to $V_t$ the following 15 variables: (1) Real Gross Domestic Product; (2) Real Personal Consumption Expenditures, (3) Industrial Production: Final Products (Market Group), (4) IP: Consumer goods, (5) Total Nonfarm Payrolls: All Employees, (6) Housing Starts: Total: New Privately Owned Housing Units Started, (7) Mfrs new orders durable goods industries, (8) Gross Domestic Product: Chain-type Price Index, (9) Average Hourly Earnings: Total Private Industries Defl by PCE(LFE) Def, (10) 3-Month Treasury Bill: Secondary Market Rate, (11) St. Louis Adjusted Monetary Base, (12) S$\&$P 500, (13) Common Stock Prices: Dow Jones Industrial Average, (14) Consumer expectations NSA, and (15) World Oil Production.

\subsection{Results}
Table~\ref{table:dfmMSErelativetoLP} shows the RMSE of the IRFs estimated by each method relative to the RMSE of the IRFs estimated by RSLP. Values above one favour RSLP. These values are reported for the experiments with the strictly and conditionally exogenous instruments.  
Section~\ref{sec:MCbenchmarks} discusses these results in more detail.

\begin{table}[h!]
	\caption{RMSE relative to RSLP in the DFM Monte Carlo experiments}\label{table:dfmMSErelativetoLP}
    \begin{tabular}{lrrrrrrrrrr}
    \toprule\toprule
          & \multicolumn{5}{c}{Strict Instrument} & \multicolumn{5}{c}{Conditional Instrument} \\
          \cmidrule(lr){2-6}\cmidrule(lr){7-11}
          & \multicolumn{1}{l}{Fed} & \multicolumn{1}{l}{IP} & \multicolumn{1}{l}{UNE} & \multicolumn{1}{l}{CPI} & \multicolumn{1}{l}{EBP} & \multicolumn{1}{l}{Fed} & \multicolumn{1}{l}{IP} & \multicolumn{1}{l}{UNE} & \multicolumn{1}{l}{CPI} & \multicolumn{1}{l}{EBP} \\
          \midrule
          & \multicolumn{10}{c}{LP methods} \\
          \midrule
  FALP  & 1.079 & 1.049 & 1.091 & 0.960 & 1.241 & 1.850 & 1.534 & 1.539 & 0.962 & 1.229 \\
    LASSO & 1.360 & 1.248 & 1.307 & 1.109 & 1.522 & 2.509 & 2.685 & 2.937 & 1.756 & 1.863 \\
    OCMT  & 3.771 & 3.268 & 3.424 & 2.641 & 4.429 & 3.327 & 3.700 & 4.359 & 3.018 & 2.504 \\
    BLP   & 0.453 & 0.753 & 0.915 & 0.436 & 0.569 & 0.870 & 1.689 & 2.007 & 0.587 & 0.976 \\
    Base  & 3.770 & 3.268 & 3.424 & 2.641 & 4.429 & 3.452 & 3.909 & 4.580 & 3.134 & 2.589 \\
    \midrule
          & \multicolumn{10}{c}{VAR methods} \\
          \midrule
    FAVAR & 0.501 & 0.653 & 0.686 & 0.499 & 0.546 & 0.464 & 0.608 & 0.625 & 0.485 & 0.498 \\
    Small  & 0.458 & 0.781 & 0.949 & 0.429 & 0.578 & 0.872 & 1.700 & 2.023 & 0.574 & 0.978 \\
    Medium & 0.484 & 0.738 & 0.831 & 0.518 & 0.480 & 0.445 & 0.773 & 0.841 & 0.501 & 0.517 \\
    \bottomrule \bottomrule
    \end{tabular}%
    \footnotesize\\
\emph{Note: This table reports the root mean squared error (RMSE) relative to RSLP for the DFM experiments with a strictly and conditionally exogenous instrument (\eqref{eq:z1_dfm} and \eqref{eq:z2_dfm}, respectively).  See text and note to Table \ref{table:fiscalMSErelativetoLP} for further details of models estimated. The DGP is parameterized by estimating the DFM on U.S. macroeconomic data. The relative RMSEs are reported for the following, representing variables which we generated from the DGP: Fed - Federal funds rate, IP - Industrial production, Unemp - Unemployment, CPI - CPI inflation, EBP - excess bond premium as per \cite{gilchrist2012credit}.}
\end{table}

\clearpage
\section{Additional results baseline Monte Carlo experiments}\label{A:MCadd}

\begin{table}[h!]
\footnotesize
	\caption{RMSE of FAVAR relative to RSLP in the baseline Monte Carlo experiments}\label{tab:horizons}
    \begin{tabular}{llrrrrrrrrrrrr}
\toprule \toprule
      &    & \multicolumn{4}{c}{Strict Instrument} & \multicolumn{4}{c}{Conditional Instrument} & \multicolumn{4}{c}{SVAR} \\
           \cmidrule(lr){3-6}\cmidrule(lr){7-10}\cmidrule(lr){11-14}
       &   & \multicolumn{2}{c}{Strong} & \multicolumn{2}{c}{Weak} & \multicolumn{2}{c}{Strong} & \multicolumn{2}{c}{Weak} & \multicolumn{2}{c}{Strong} & \multicolumn{2}{c}{Weak} \\
           \cmidrule(lr){3-4}\cmidrule(lr){5-6} \cmidrule(lr){7-8}\cmidrule(lr){9-10}\cmidrule(lr){11-12}\cmidrule(lr){13-14}
      h  & method  & \multicolumn{1}{l}{Tax} & \multicolumn{1}{l}{Cap } & \multicolumn{1}{l}{Tax} & \multicolumn{1}{l}{Cap } & \multicolumn{1}{l}{Tax} & \multicolumn{1}{l}{Cap } & \multicolumn{1}{l}{Tax} & \multicolumn{1}{l}{Cap } & \multicolumn{1}{l}{Tax} & \multicolumn{1}{l}{Cap } & \multicolumn{1}{l}{Tax} & \multicolumn{1}{l}{Cap } \\
          \midrule
    0     & Base  & 13.919 & 16.408 & 12.175 & 14.143 & 38.399 & 3.432 & 5.509 & 2.757 & 130.860 & 2.961 & 11.985 & 2.248 \\
    0     & FALP  & 0.956 & 0.952 & 1.048 & 1.048 & 1.794 & 0.961 & 1.534 & 1.104 & 5.578 & 1.054 & 2.998 & 1.282 \\
    0     & FAVAR & 8.469 & 9.182 & 7.384 & 7.629 & 1.048 & 0.983 & 1.875 & 2.507 & 2.509 & 1.811 & 1.959 & 4.649 \\ \midrule
    1     & Base  & 13.764 & 12.699 & 11.641 & 10.796 & 22.921 & 1.902 & 4.644 & 1.673 & 4.525 & 1.822 & 0.703 & 1.641 \\
    1     & FALP  & 0.973 & 0.942 & 1.055 & 1.035 & 1.018 & 0.968 & 1.315 & 1.048 & 0.647 & 0.979 & 0.716 & 1.082 \\
    1     & FAVAR & 2.752 & 2.187 & 2.303 & 1.826 & 0.665 & 0.941 & 2.944 & 1.473 & 1.164 & 0.956 & 7.302 & 1.865 \\ \midrule
    2     & Base  &  & 24.520 &  & 20.548 &  & 1.608 &  & 1.455 &  & 1.160 &  & 1.082 \\
    2     & FALP  &  & 0.940 &  & 1.034 &  & 0.972 &  & 1.050 &  & 0.933 &  & 1.073 \\
    2     & FAVAR &  & 1.314 &  & 1.141 &  & 0.920 &  & 1.124 &  & 0.912 &  & 1.078 \\ \midrule
    3     & Base  & 6.298 & 16.246 & 5.452 & 13.893 & 1.498 & 1.476 & 1.391 & 1.350 & 0.946 & 0.967 & 0.887 & 0.890 \\
    3     & FALP  & 0.944 & 0.968 & 1.047 & 1.048 & 0.955 & 0.943 & 1.013 & 0.992 & 0.935 & 0.923 & 1.126 & 1.009 \\
    3     & FAVAR & 0.527 & 0.670 & 0.448 & 0.577 & 0.904 & 0.898 & 1.124 & 0.968 & 0.907 & 0.878 & 1.056 & 0.984 \\ \midrule
    4     & Base  & 10.000 & 15.213 & 8.536 & 13.057 & 1.500 & 1.492 & 1.415 & 1.367 & 0.883 & 0.922 & 0.830 & 0.849 \\
    4     & FALP  & 0.955 & 0.942 & 1.039 & 1.026 & 0.950 & 0.951 & 1.017 & 1.009 & 0.934 & 0.934 & 1.068 & 1.059 \\
    4     & FAVAR & 0.291 & 0.372 & 0.238 & 0.308 & 0.900 & 0.874 & 0.951 & 0.899 & 0.901 & 0.862 & 0.965 & 0.874 \\ \midrule
    5     & Base  & 4.982 & 12.747 & 4.350 & 11.135 & 1.442 & 1.505 & 1.357 & 1.364 & 0.871 & 0.934 & 0.813 & 0.865 \\
    5     & FALP  & 0.956 & 0.947 & 1.052 & 1.048 & 0.954 & 0.936 & 1.021 & 0.988 & 0.923 & 0.917 & 1.045 & 1.029 \\
    5     & FAVAR & 0.123 & 0.184 & 0.103 & 0.157 & 0.874 & 0.400 & 0.841 & 0.417 & 0.883 & 0.400 & 0.823 & 0.417 \\ \midrule
    6     & Base  & 6.114 & 22.655 & 5.302 & 19.762 & 1.440 & 1.490 & 1.361 & 1.363 & 0.968 & 0.927 & 0.905 & 0.858 \\
    6     & FALP  & 0.948 & 0.949 & 1.038 & 1.036 & 0.934 & 0.950 & 0.999 & 1.027 & 0.956 & 0.915 & 1.111 & 1.010 \\
    6     & FAVAR & 0.072 & 0.099 & 0.059 & 0.085 & 0.852 & 0.197 & 0.750 & 0.208 & 0.860 & 0.199 & 0.756 & 0.211 \\
\bottomrule \bottomrule
    \end{tabular}%
    \footnotesize\\
\emph{Note: This table reports the root mean squared error (RMSE) of FAVAR relative to RSLP for the IV and SVAR identification under both the strong and weak factor structure settings. Strict and conditional instruments refer to instruments generated by \eqref{eq:z_exo} and \eqref{eq:z_endo}, respectively.  Since the IRF of the tax rate is scaled to a unit tax shock at horizon $h=2$, the corresponding RMSEs are zero and the relative RMSE is undefined.
}
\end{table}%

\section{Implementation details empirical applications}\label{A:benchmark_apps}
The methods are implemented as outlined in Appendix~\ref{A:benchmarks}, where the macroeconomic time series included in $G_t$ for RSLP are considered as information series $\{y^*_s\}_{s=1}^t$. We include three principal components in FALP and FAVAR in the technology shock application and four in the monetary policy shock application, as suggested by the Granger causality test with the variables in $V_t$ proposed by \citet{forni2014sufficient}. The BVAR includes the same variables as in $V_t$ and considers four lags in the technology shock application and twelve in the monetary policy shock application. 

The medium BVAR includes in total 20 variables, which means that in addition to $V_t$ another 18 variables in the technology shock application and 16 in the monetary policy shock application have to be selected. For the technology shock application, the following variables are selected from FRED-QD (FRED mnemonics in parentheses): 
(1) Consumer Price Index for All Urban Consumers: All Items (CPI), (2) Industrial Production Index (IP), (3) Effective Federal Funds Rate (FEDFUNDS), (4) Help-Wanted Index (HWIx), (5) Producer Price Index: Commodities: Metals and metal products (PPICMM), (6) S$\&$P’s Common Stock Price Index: Composite (S$\&$P 500), (7) S$\&$P’s Common Stock Price Index: Industrials (S$\&$P:indust), (8) Housing Starts: Total (HOUST), (9) New Private Housing Units Authorized by Building Permits (PERMIT), (10) 1-Year Treasury Constant Maturity Rate (GS1), (11) Real M1 Money Stock (M1REAL), (12) Real M2 Money Stock  (M2REAL), (13) CP/GDPDEF, (14) GDPDEF, (15) dtfp, (16) dtfp-util, (17) Business Condition next year, (18) Expected Index. 

For the monetary policy shock application, the following variables are selected from FRED-MD:
(1) Civilian Unemployment Rate (UNRATE), (2) IP: Final Products (Market Group) (IPFINAL), (3) IP: Consumer Goods (IPCONDGD), (4) Initial Claims (CLAIMSx), (5) All Employees: Total nonfarm (PAYEMS), (6) Total Business Inventories (BUSINVx), (7) Total Business: Inventories to Sales Ratio (ISRATIOx), (8) CPI : Apparel (CPIAPPSL), (9) Avg Hourly Earnings: Construction (CES2000000008), (10) Consumer Sentiment Index (UMCSENTx), (11) Housing Starts: Total New Privately Owned (HOUST), (12)New Private Housing Permits (PERMIT), (13) Nonrevolving consumer credit to Personal Income (CONSPI), (14) S$\&$P’s Common Stock Price Index: Composite (S$\&$P 500), (15) 3-Month Treasury Bill (TB3MS), (16) Japan / U.S. Foreign Exchange Rate (EXJPUSx).

 \end{document}